\documentclass[a4paper, 10pt,aps,preprint,superscriptaddress,nofootinbib,floatfix,pre]{revtex4-1}
\usepackage[pdftex]{graphicx}
\usepackage{latexsym,amsmath,verbatim}
\usepackage{xcolor}
\usepackage[hidelinks,unicode=true]{hyperref}
\hypersetup{colorlinks=true,linkcolor=blue,urlcolor=blue,citecolor=blue,pdfhighlight=/N
}
\usepackage{longtable}

\usepackage{lipsum}
\usepackage[english]{babel}
\usepackage{microtype}
\usepackage[normalem]{ulem}

\newcommand{\av}[1]{\left\langle {#1} \right\rangle}
\newcommand{\kmax}{k_\mathrm{max}}
\newcommand{\be}{\begin{equation}}
\newcommand{\ee}{\end{equation}}

\newcommand{\change}[1]{#1}

\renewcommand*{\arraystretch}{1.3}

\begin{document}

\title{The localization of non-backtracking centrality in networks \\ and its
physical consequences}

\author{Romualdo Pastor-Satorras}
     
\affiliation{Departament de F\'{\i}sica, Universitat Polit\`ecnica de
Catalunya, Campus Nord B4, 08034 Barcelona, Spain}

\author{Claudio Castellano} 

\affiliation{Istituto dei Sistemi Complessi (ISC-CNR), Via dei Taurini 19,
I-00185 Roma, Italy\\ \ \\
Corresponding author: romualdo.pastor@upc.edu}

\begin{abstract}
  The spectrum of the non-backtracking matrix plays a crucial role in
  determining various structural and dynamical properties of networked
  systems, ranging from the threshold in bond percolation and non-recurrent
  epidemic processes, to community structure, to node importance.  Here we
  calculate the largest eigenvalue of the non-backtracking matrix and the
  associated non-backtracking centrality for uncorrelated random networks,
  finding expressions in excellent agreement with numerical results. We show
  however that the same formulas do not work well for many real-world
  networks.  We identify the mechanism responsible for this violation in the
  localization of the non-backtracking centrality on network subgraphs whose
  formation is highly unlikely in uncorrelated networks, but rather common
  in real-world structures.  Exploiting this knowledge we present an
  heuristic generalized formula for the largest eigenvalue, which is
  remarkably accurate for all networks of a large empirical dataset.  We
  show that this newly uncovered localization phenomenon allows to
  understand the failure of the message-passing prediction for the
  percolation threshold in many real-world structures.
\end{abstract}

\maketitle

\section{Introduction}

The non-backtracking (NB) operator is a binary matricial representation of
the topology of a network, whose elements represent the presence of
\textit{non-backtracking} paths between pairs of different nodes, traversing
a third intermediate one\change{~\cite{hashimoto1989,Martin2014}}.
By means of a message-passing
approach~\cite{10.5555/1592967}, the NB matrix finds a natural use in the
representation of dynamical processes on networks, such as
percolation~\cite{Cohen00,Callaway2000} and non-recurrent
epidemics~\cite{PastorSatorras2015}, where a
spreading process cannot affect twice a given node, and therefore
backtracking propagation paths are inhibited~\cite{Karrer2014,Karrer2010}.
Within this approach, the bond percolation threshold and the epidemic
threshold in the SIR model~\cite{PastorSatorras2015} are found to be inversely
proportional to the largest eigenvalue (LEV) of the NB matrix, $\mu_M$.
\change{The spectrum of the non-backtracking matrix is relevant
  also for other problems in network science, such as
  community structure~\cite{Krzakala2013} and node
  importance~\cite{Martin2014,Morone2015,Radicchi2016,Torres2020}.}

The principal eigenvector (PEV) associated to the LEV of the NB matrix has
been recently used to build a new measure of node importance or
centrality~\cite{Newman10}.  A classical measure of node centrality is given
by eigenvector centrality, based on the idea that a node is central if it is
connected to other central nodes.  In this perspective, eigenvector
centrality of node $i$ is defined as the $i$-th component of the principal
eigenvector of the adjacency matrix~\cite{Bonacich72}. Eigenvector
centrality has the drawback of being strongly affected by the presence of
large hubs, which exhibit an exceedingly large component of the adjacency
matrix PEV because of a peculiar self-reinforcing bootstrap effect.  The hub
is highly central since it has a large number of mildly central neighbors;
the neighbors are in their turn central just because of their vicinity with
the highly central hub~\cite{Goltsev2012,Martin2014}.  In terms of the
adjacency matrix this self-reinforcement is revealed by the localization of
the PEV on a star graph composed by the largest hub and its immediate
neighbors.  To correct for this feature, in Ref.~\cite{Martin2014} it was
proposed to build a centrality measure using the NB matrix, in such a way as
to avoid backtracking paths that could artificially inflate a hub's
centrality.  In this way, an alternative non-backtracking centrality (NBC)
of nodes was defined, in which the effect of hubs is strongly suppressed.

Consider an unweighted undirected complex network with $N$ nodes and $E$
edges.  The non-backtracking (NB) matrix $\mathbf{B}$ is a representation of
the network topology in terms of a $2E \times 2E$ non-symmetric matrix in
which rows and columns represent virtual directed edges $j \to i$ pointing
from node $j$ to node $i$, taking the value
\begin{equation}
  \label{eq:6}
  B_{j \to i, m \to \ell} = \delta_{j \ell} ( 1- \delta_{i m}),
\end{equation}
where $\delta_{i j}$ represents the Kronecker symbol. Each NB matrix element
represents a possible walk in the network composed by a pair of directed
edges, one pointing from node $m$ to node $\ell$, and the other from node
$j$ to node $i$. The element is nonzero when the edges share the central
node ($j = \ell$), and when the walk does not return to the first node ($m
\neq i$). 

The principal eigenvector  $v_{j \to i}$ of the NB matrix, associated to the
largest eigenvalue (LEV) $\mu_M$, is given by the relation 
\begin{equation}
  \label{eq3}
  \mu_M v_{j \to i} = \sum_{m \to l}  B_{j \to i, m \to l} v_{m \to l}.
\end{equation}
Since $\mathbf{B}$ is a non-negative matrix, the Perron-Frobenius
theorem~\cite{Gantmacher} guarantees that $\mu_M$ and all components $v_{j
  \to i}$ are positive, provided that the matrix is irreducible.

The element $v_{j \to i}$ expresses the centrality of node $j$, disregarding
the possible contribution of node $i$. The non-backtracking centrality
$x_i$ of node $i$ is defined as~\cite{Martin2014}
\begin{equation}
  \label{eq:10}
  x_i = \sum_j A_{ij} v_{j \to i},
\end{equation}
where $A_{ij}$ is the network adjacency matrix.  If the PEV of the NB matrix
is normalized as $\sum_{j \to i} v_{j \to i} = \sum_{j, i} A_{ji} v_{j \to i}
= 1$, which is valid if $\mathbf{B}$ is irreducible, then the natural
normalization $\sum_i x_i = 1$ emerges. 

\section*{Results}
\subsection*{Theory for uncorrelated random networks}
\label{sec:theory}

The NBC can be practically calculated by using the Ihara-Bass
determinant formula~\cite{bass1992,Martin2014}, which shows that the NBC
values $x_i$ correspond to the first $N$ elements of the PEV of the $2N
\times 2N$ matrix
\begin{equation}
  \label{eq:9}
  \mathbf{M} = \left(
    \begin{array}{cc}
      \mathbf{A} & \mathbf{I} - \mathbf{D}\\
      \mathbf{I} & \mathbf{0}
    \end{array}
  \right),
\end{equation}
where $\mathbf{A}$ is the adjacency matrix, $\mathbf{I}$ is the identity
matrix, and  $\mathbf{D}$ is a diagonal matrix of elements $D_{ij} =
\delta_{ij} k_i$.  Using the Ihara-Bass formalism~\cite{PhysRevE.91.010801}
(see Method~\ref{RadicchiMartin}) one can express, in full generality, the
leading eigenvalue $\mu_M$ in terms of the NBC as
\begin{equation}
  \mu_M = \frac{\sum_i k_i x_i}{\sum_i x_i} -1.
  \label{eq:mu_1_exact}
\end{equation}

Following Ref.~\cite{Martin2014} (see Method~\ref{RadicchiMartin}), it is
possible to argue that, for uncorrelated random networks,
\change{i.e., networks with a given degree sequence but completely random in
all other respects~\cite{Newman10}},
the dependence of
the components of the NB matrix PEV is 
\begin{equation}
   v_{j \to i} \sim k_j-1 .
\end{equation}
Introducing this relation into the definition of the NBC, Eq.~\eqref{eq:10},
and applying the normalization $\sum_i x_i = 1$, we obtain 
\begin{equation}
  x_i^\mathrm{un} = \frac{\sum_j A_{ij} (k_j - 1)}{\sum_j k_j(k_j - 1)},
  \label{eq:nbc_approximation}
\end{equation}
that, inserted into Eq.~\eqref{eq:mu_1_exact}, leads to
\begin{equation}
  \mu_M^\mathrm{un} =  \frac{\sum_{ij} (k_i - 1) A_{ij} (k_j -1)}{\sum_j k_j(k_j - 1)}.
  \label{eq:lev_approximation}
\end{equation}

These expressions constitute an improvement over previous
results~\cite{Krzakala2013,Martin2014,PhysRevE.91.010801},
namely
\begin{equation}
  x_i^\mathrm{an} = \frac{k_i}{\av{k}N}, \quad
  \mathrm{and} \quad \mu_M^\mathrm{an} = 
  \frac{\av{k^2}}{\av{k}} - 1,   
  \label{eq:an_results}
 \end{equation}
($\av{k^n}$ is the $n$-th moment of the degree
distribution), which can be recovered from
Eqs.~(\ref{eq:nbc_approximation})
and~(\ref{eq:lev_approximation}) by replacing the network
adjacency matrix with its annealed approximated
value $\bar{A}_{ij} = k_i k_j/(\av{k}N)$~\cite{Dorogovtsev2008,Boguna09}.

\subsection*{Test on synthetic networks}
\label{sec:synthetic}

We now check the predictions developed above with
the LEV $\mu_M$ and the NBC $x_i$ determined numerically by applying
the power iteration method~\cite{golub2012matrix} to the
Ihara-Bass matrix $\mathbf{M}$ for random uncorrelated networks
with a power-law degree distribution $P(k) \sim k^{-\gamma}$,
generated using the uncorrelated
configuration model (UCM)~\cite{Catanzaro2005}.
\begin{figure}[t]
  \centering \includegraphics[width=0.7\columnwidth]{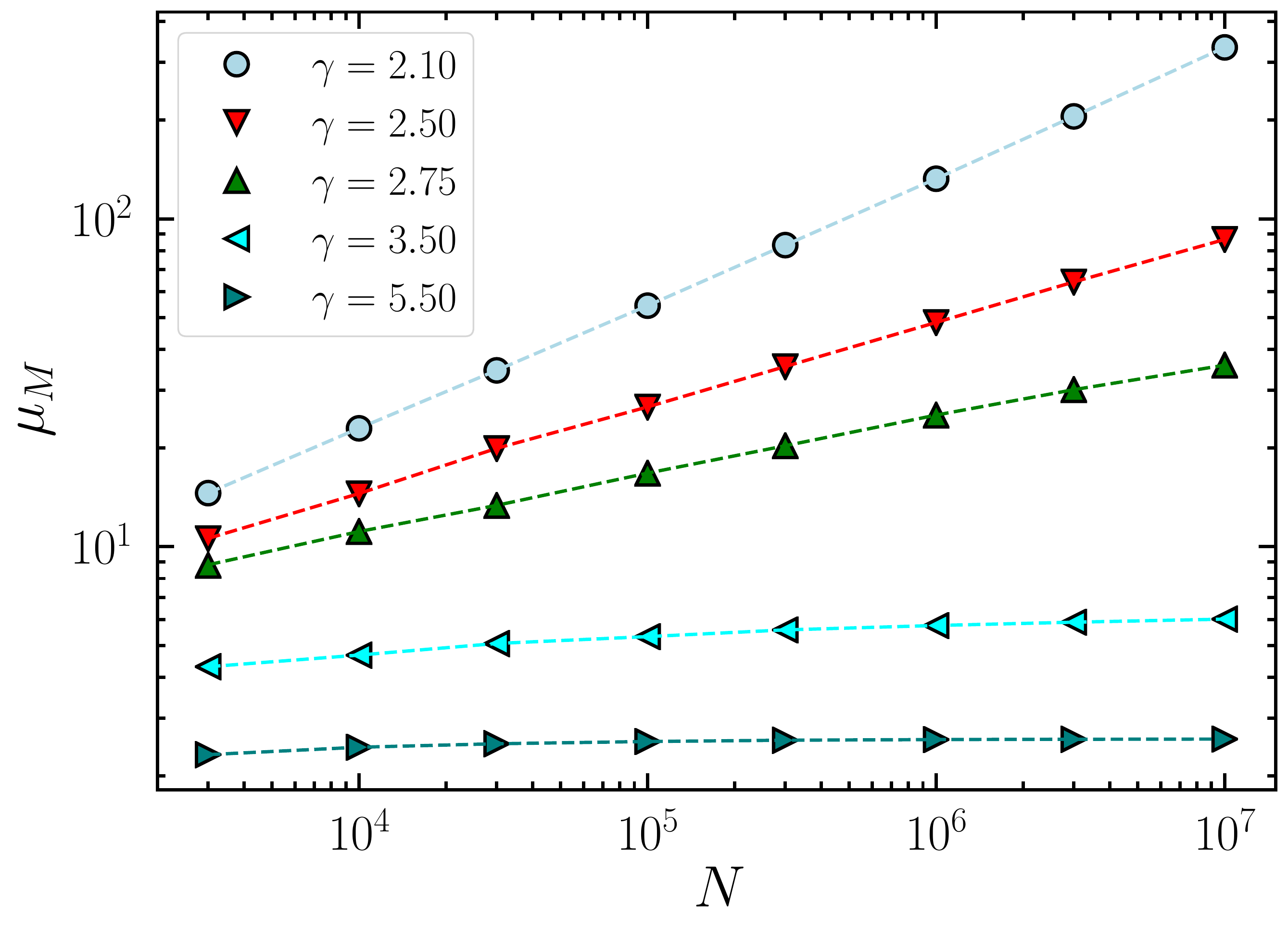}
  \caption{{\bf $\mu_M$ for uncorrelated networks.}
    Scaling of the LEV of the NB matrix, $\mu_M$, as a function of
    network size $N$ in power law UCM networks with different degree
    exponent $\gamma$. Dashed lines correspond to the theoretical prediction
  Eq.~\eqref{eq:lev_approximation}. Simulations results correspond to the
average over $25$ different network realizations. Error bars are smaller
than symbols size.}
  \label{fig:hashimoto_lev_ucm}
\end{figure}
\begin{figure*}[t]
  \centering \includegraphics[width=0.9\textwidth]{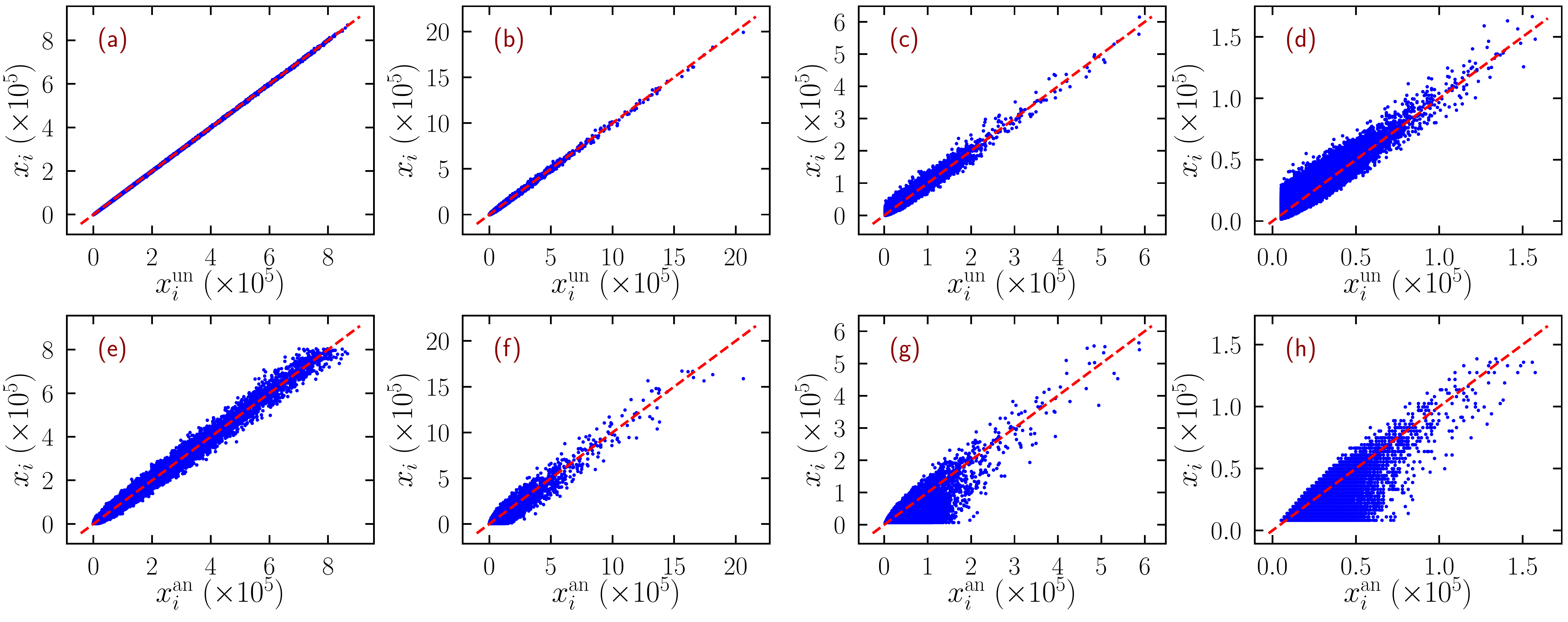}
  \caption{{\bf NBC for uncorrelated networks.}
    Scatter plot of the numerical NBC $x_i$ in power-law UCM networks
    of size $N=10^6$ with different degree exponent $\gamma$, as a function
    of the theoretical predictions $x_i^\mathrm{un}$ in
    Eq.~\eqref{eq:nbc_approximation} (top row) and  $x_i^\mathrm{an}$ in
    Eq.~\eqref{eq:an_results} (bottom row). The dashed lines represent the
    curve $y = x$. Degree exponents considered are $\gamma = 2.10$ (a) and
  (e); $\gamma = 2.75$ (b) and (f); $\gamma = 3.50$ (c) and (g); $\gamma =
4.50$ (d) and (h).}
  \label{fig:hashimoto_pev_ucm}
\end{figure*}
In Fig~\ref{fig:hashimoto_lev_ucm} we present, as a function of the network
size $N$, a comparison between the NB LEV, $\mu_M$, evaluated numerically
and our theoretical prediction Eq.~\eqref{eq:lev_approximation}.  The match
between theory and simulation is excellent.  However, also
Eq.~(\ref{eq:an_results}) gives very accurate results, differing in average
by less than $0.5\%$ from the theoretical result
Eq.~\eqref{eq:lev_approximation}. A much more noticeable improvement is
observed instead for the NB centrality $x_i$, for which annealed network
approximation does not provide accurate predictions (see
Fig.~\ref{fig:hashimoto_pev_ucm}, bottom row).  In
Fig.~\ref{fig:hashimoto_pev_ucm} (top row) we show the dependence of the NBC
$x_i$ on the structure of the adjacency matrix, as given by
Eq.~\eqref{eq:nbc_approximation}, namely $x_i \sim \sum_j A_{ij} (k_j -1)$.
The analytical expression is extremely accurate for values of $\gamma < 3$.
For $\gamma > 3$, although some scattering can be observed with respect to
the expected value, the prediction is still good, much more accurate than
the annealed network approximation.  More evidence about the superior
accuracy of our approach is found considering the inverse participation
ratio $Y_4(N)$ as a function of network size (see
Method~\ref{appendix:localization}).

\subsection*{Non-backtracking principal eigenvalue of characteristic
subgraphs}
\label{sec:motifs}

The non-backtracking centrality was introduced with the goal of overcoming
the flaws of eigenvector centrality, due to the localization of the
adjacency matrix principal eigenvector on star graphs surrounding hubs of
large degree, that artificially inflate their own eigenvector
centrality~\cite{Martin2014}.
For the NBC the addition of a large hub to an
otherwise homogeneous network has a limited impact.
Indeed, the addition of
a dangling hub of degree $K$, connected to $K-1$ leaves of degree $1$ and to
a generic network by a single edge, does not alter at all the value of
$\mu_M$~\cite{Krzakala2013,Martin2014} (see Method~\ref{appendix:isolated}).
In the case of a hub integrated into the
network, connected to $K$ other random nodes in the graph,
Ref.~\cite{Martin2014} argued, from the perspective of the annealed network
approximation, that its effect is irrelevant in the thermodynamic limit. A
more elaborate analysis (see Method~\ref{appendix:isolated})
shows that this is true unless $K \gg (N/\av{k})^{1/2}$. Only in this
case an integrated hub has an effect and leads to a PEV significantly
larger than the PEV of the original network and 
scaling as $[\av{k} K(K-1)/N]^{1/3}$.

However, it is possible that other types of subgraphs play for the NB
centrality the same role that star graphs play for eigenvector centrality:
They can have, alone, large values of $\mu_M$, so that, if present within an
otherwise random network, they determine $\mu_M$ of the whole structure, with
the overall NBC localized on them.
We now show that these subgraphs actually exist and can have dramatic effects.

As noticed in Ref.~\cite{Martin2014}, the simplest example is a clique of
size $K_c$, which is associated to $\mu_M^\mathrm{clique}=K_c-2$.  If $K_c$
is large enough, $\mu_M^\mathrm{clique}$ can dominate over $\mu_M^\mathrm{un}$.
But also a homogeneous (Poisson) subgraph of average degree $\av{k}$, for
which $\mu_M=\av{k}$~\cite{Martin2014,Krzakala2013}, can become the
substrate of a localized NB PEV if $\av{k}$ is sufficiently large.

Apart from these simple examples, a less trivial one is the case of
\textit{overlapping hubs}, i.e., a set of $n$ hubs of degree $K$,  connected
to the same $K$ leaves of degree $n$, see Supplementary
Figure~SF1. The intrinsic LEV associated to such a
structure is (see Method~\ref{appendix:isolated})
\begin{equation}
  \mu_M^\mathrm{oh} = \sqrt{(n-1)(K-1)}.
  \label{eq:muoh}
\end{equation}
This last case is particularly important, since $\mu_M^\mathrm{oh}$ can
become very large due to a few overlapping hubs of very large degree $K$, or
due to a large number of hubs with moderate overlap $K$.

\subsection*{Localization in real-world networks}
\label{sec:real}

\begin{figure*}
  \centering \includegraphics[width=0.9\textwidth]{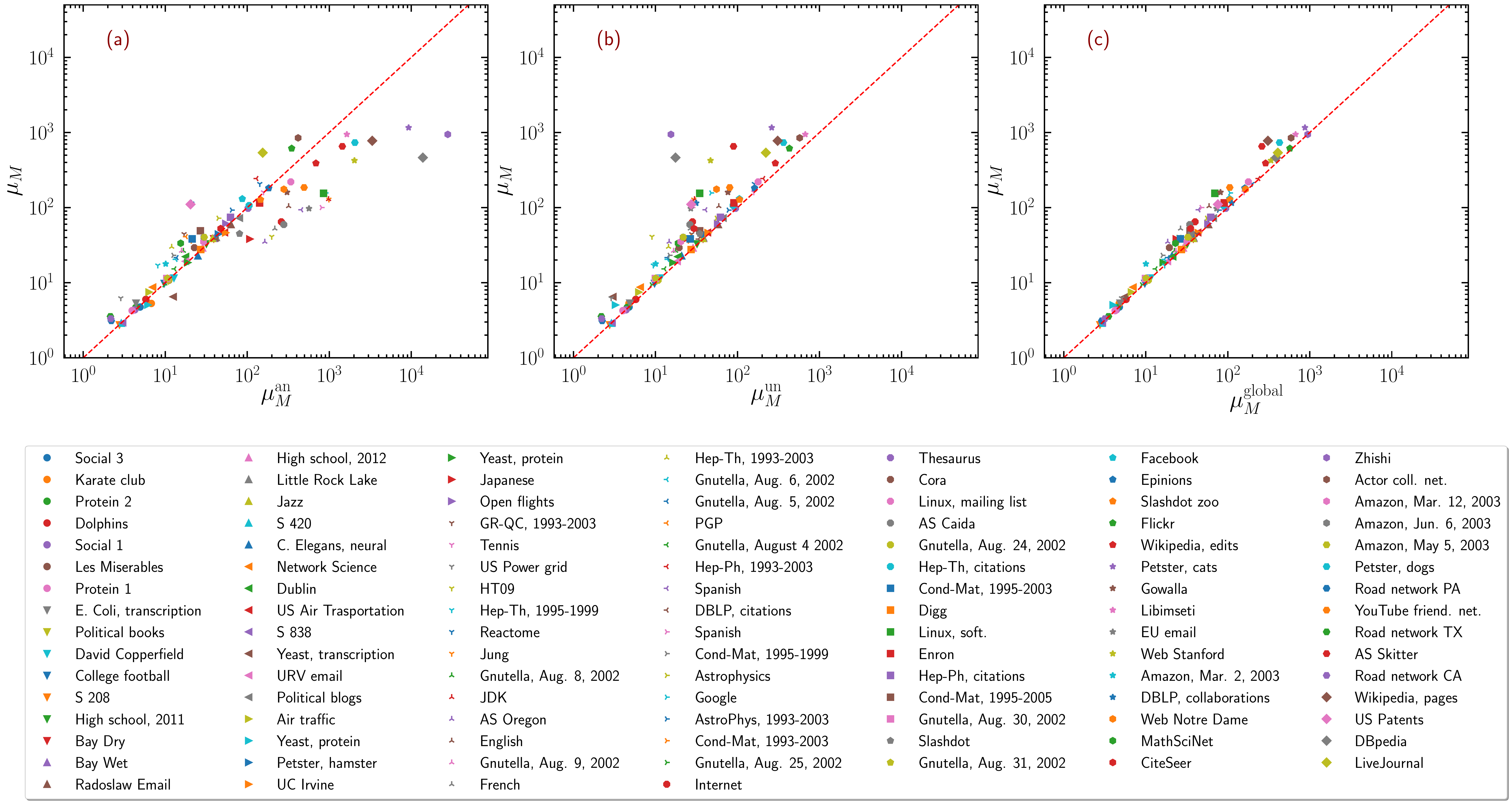}
  \caption{{\bf Test of theoretical approaches for real-world networks.}
    LEV of the NB matrix, $\mu_M$, as a function of the theoretical
    predictions $\mu_M^\mathrm{\change{un}}$ [Eq.~\eqref{eq:lev_approximation}] (a)
    $\mu_M^\mathrm{\change{an}}$ [Eq.~\eqref{eq:an_results}] (b), and
  $\mu_M^\mathrm{global}$ [Eq.~(\ref{muprediction})] (c), for the set of 109
real-world networks described in Supplementary Table ST1.}
 \label{fig:lev_real}
\end{figure*}

In Fig.~\ref{fig:lev_real}(a) and~\ref{fig:lev_real}(b) we compare the
theoretical predictions derived for uncorrelated
and annealed networks with the values of $\mu_M$ computed numerically for a
set of 109 real-world networks of diverse origin (see Supplementary
Table ST1 for details).  In opposite ways, both predictions,
$\mu_M^\mathrm{an}$ and $\mu_M^\mathrm{un}$, fail to provide an accurate
approximation of empirical results for many networks. In the
most noticeable cases, the networks \texttt{Zhishi} and \texttt{DBpedia},
the uncorrelated prediction Eq.~\eqref{eq:lev_approximation} largely
underestimates the value of $\mu_M$, while the annealed network prediction
Eq.~\eqref{eq:an_results} largely overestimates it.

To shed light on the origin of these discrepancies, in Supplementary
Figure~SF2 we compare the empirical NBC, $x_i$,
with the theoretical prediction $x_i^\mathrm{un}$ for four real-world networks in
which the predictions largely fail.  We observe that, in all networks, a
few nodes assume an exceedingly large value of $x_i$, i.e., the NBC is
localized on a very small subset of nodes, which includes the largest hubs.

It is clear that, in order to obtain an accurate prediction of $\mu_M$ in
real-world networks, it is necessary to take into account the possible
localization of the NB centrality on subgraphs which, despite being
relatively small, may determine $\mu_M$ for the whole structure.  In
previous paragraphs, we have seen that two special subgraphs, a large
clique/relatively dense homogeneous graph, or a set of overlapping hubs, may
become the set where NBC gets localized if the associated $\mu_M$ is larger
than the one for the rest of the network.  It is then natural to postulate
(in analogy with what happens for the adjacency
matrix~\cite{Castellano2017}) that the overall $\mu_M$ is well approximated
by the maximum among Eq.~(\ref{eq:lev_approximation}) and the $\mu_M^{(s)}$
values associated to each possible network subgraph $s$~\footnote{
  \change{We note here that, while in the case of the adjacency matrix this
    result is exact due to the Rayleigh’s
    inequality~\cite{PVM_graphspectra}, for the NB matrix we simply proceed
    by analogy. As we will see later on, however, the conjecture turns out
to be quite accurate.}}.  An exhaustive search among all subgraphs is
computationally impractical.  However, if we limit ourselves to the types of
subgraphs discussed above, it is numerically easy to find reasonable
estimates of their maximum LEVs.  \change{The hubs, either dangling or
integrated, provide a negligible contribution, as we can check numerically.}
The $K$-core decomposition (see Method~\ref{appendix:kcores}) provides, as
the core with maximum index, an approximation of the densest subgraph in the
network.  The value $\mu_M^\mathrm{core}$ associated to such max $K$-core,
which can be either a clique or a relatively dense homogeneous graph, is a
good estimate of the maximum LEV among these types of subgraphs.  Concerning
$\mu_M^\mathrm{oh}$, the pair of $n$ and $K$ values maximizing
Eq.~(\ref{eq:muoh}) can be well approximated by a heuristic greedy algorithm
described in Method~\ref{appendix:heuristicalgorithm}. 

Following this line of reasoning, we can then write an approximate expression for
the NB LEV in generic networks as 
\begin{equation}
  \mu_M^\mathrm{global} = \max \left\{ \mu_M^\mathrm{un}, \,
  \mu_M^\mathrm{oh}, \, \mu_M^\mathrm{core} \right\},
  \label{muprediction}
\end{equation}
\change{where $\mu_M^\mathrm{core}$ is computed as the largest eigenvalue of
the NB matrix defined by the subgraph spanned by the maximum $K$-core.} The
comparison of Eq.~(\ref{muprediction}) with empirical results in real-world
networks, displayed in Fig.~\ref{fig:lev_real}(c), reveals a striking
accuracy in all cases and substantiates the predictive power of
Eq.~(\ref{muprediction}) for the LEV of the non-backtracking matrix on
generic real-world networks.  The spontaneous formation of large cliques or
sets of overlapping hubs is exceedingly improbable in uncorrelated networks.
A $K$-core structure exists only for $\gamma<3$~\cite{Dorogovtsev2006} but
in that case $\mu_M^\mathrm{core} \simeq \mu_M^\mathrm{un}$.  As a
consequence, for all uncorrelated networks Eq.~(\ref{muprediction}) gives
back Eq.~(\ref{eq:lev_approximation}).

\subsection*{Application to percolation}

Spectral properties of the non-backtracking matrix are at the heart of the
message-passing theory for bond percolation~\cite{Karrer2014}: For locally
tree-like networks, the percolation threshold is given by the inverse of the
NB matrix LEV,
\begin{equation}
  p_c = \frac{1}{\mu_M}.
  \label{pc}
\end{equation}

\begin{figure*}
  \centering
  \includegraphics[width=0.9\textwidth]{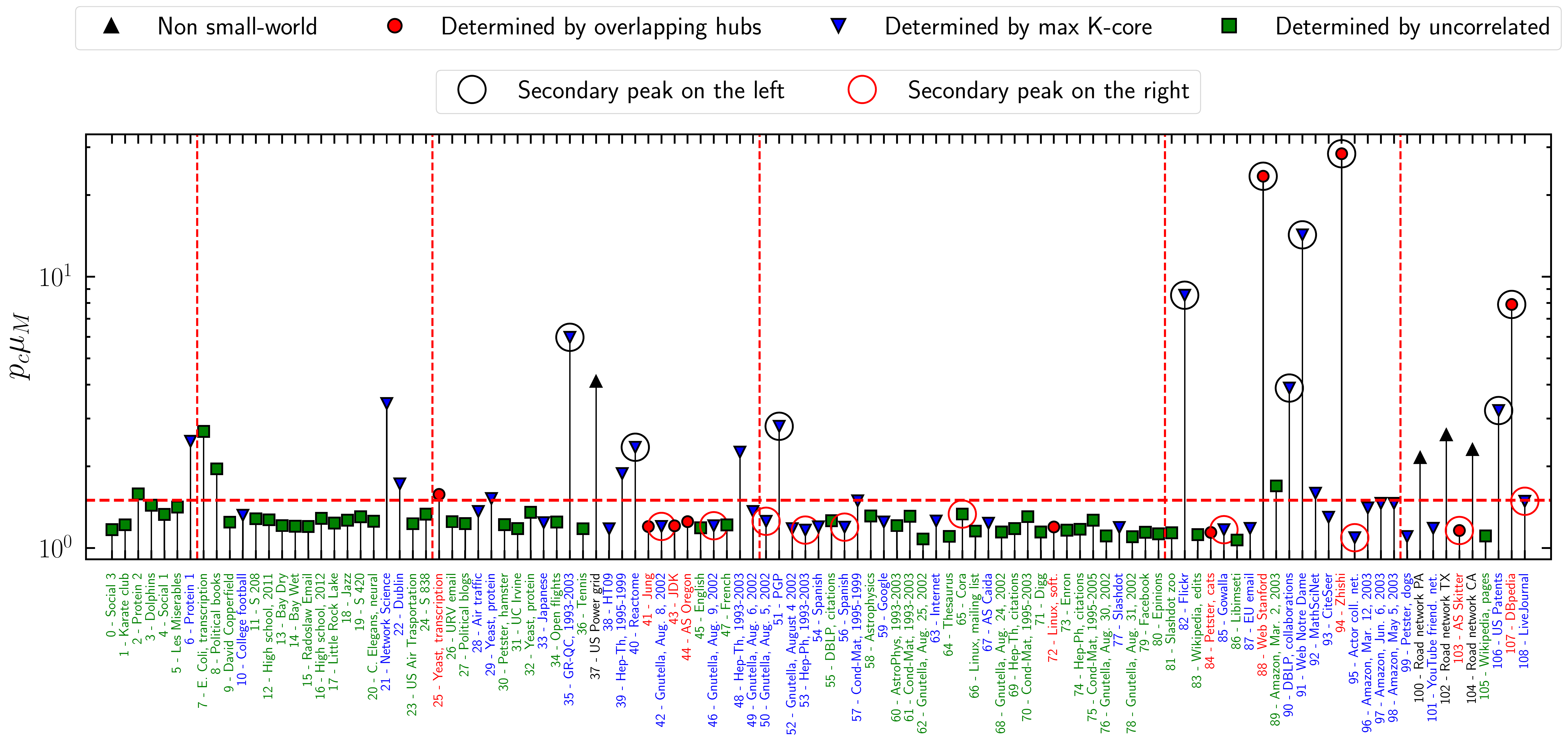}
  \caption{{\bf Test of message-passing prediction for bond percolation threshold in real-world networks.}
    The bond percolation threshold $p_c$ determined numerically from the main
    peak of the susceptibility is divided by
    the message-passing prediction [Eq.~(\ref{pc})] and plotted for the 109
    real-world networks considered.
    Below the horizontal dashed red line
    the prediction is accurate within $50\%$.  Vertical dashed lines
    represent the size scale of the networks: from left to right $N=10^2$,
    $10^3$, $10^4$, $10^5$, and $10^6$. Symbols show which of the terms in
  Eq.~(\ref{muprediction}) is maximal.  Symbols are surrounded by a black
(red) circle in case a secondary peak appears in the susceptibility on the
  left (right) of the main peak.}
  \label{fig:scatter_pc}
\end{figure*}

\begin{figure*}
  \centering
  \includegraphics[width=0.9\textwidth]{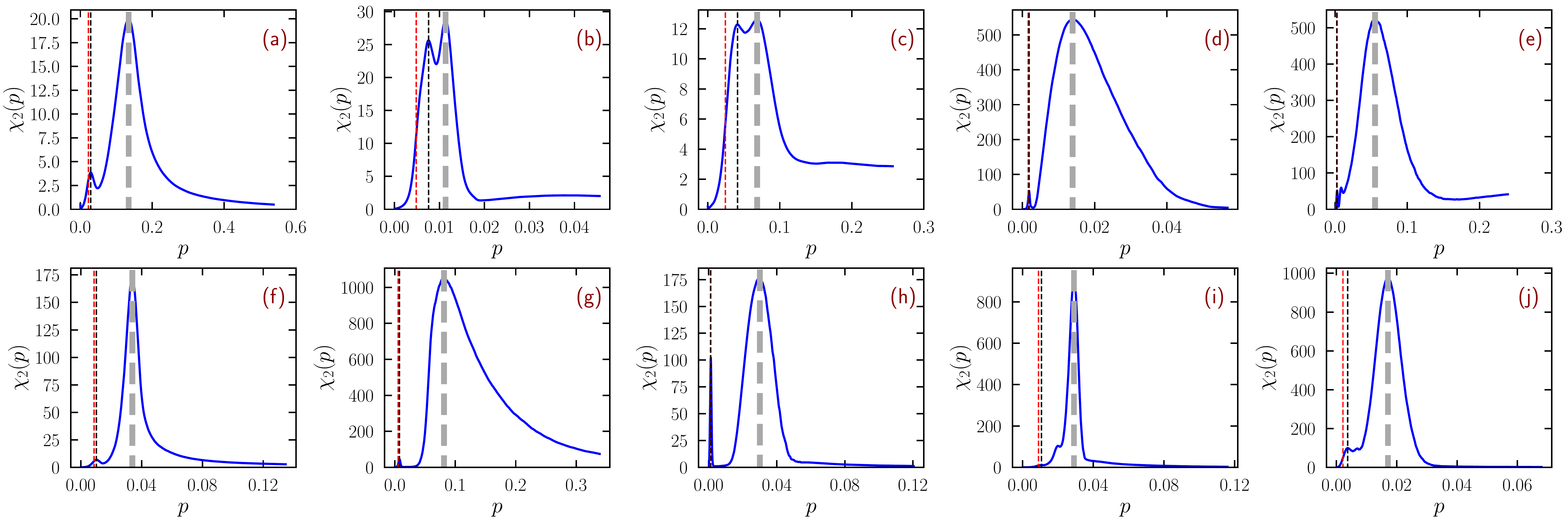}
  \caption{{\bf Susceptibility plots for networks exhibiting a secondary
    peak on the left.}
    Numerical bond percolation susceptibility for the networks (a):
    \texttt{GR-QC, 1993-2003}; (b): \texttt{Reactome}; (c): \texttt{PGP};
    (d): \texttt{Flickr}; (e): \texttt{Web Stanford}; (f): \texttt{DBLP,
    collaborations}; (g): \texttt{Web Notre Dame}; (h): \texttt{Zhishi};
    (i): \texttt{US Patents}; and (j): \texttt{DBpedia}. The global maximum
    of the susceptibility $\chi_2(p)$, indicating the percolation threshold,
    is marked by a gray vertical bar.  Black vertical lines indicate the
    position of the secondary peak.  Red vertical lines signal the value of
    the prediction $1/\mu_M$.  Notice that for three of the networks
  (\texttt{Web Stanford}, \texttt{Zhishi} and \texttt{DBpedia}) the NBC is
localized on overlapping hubs, while for the others localization occurs on
the max K-core.}
  \label{fig:susceptibility_left_peak}
\end{figure*}

A comparison of this prediction with results obtained numerically for our
set of real-world networks is presented\footnote{A similar test was already
performed in Ref.~\cite{PhysRevE.91.010801}.} in Fig.~\ref{fig:scatter_pc},
where the percolation threshold $p_c$ is obtained as  the position of the
main susceptibility peak (see Method~\ref{perc}).  In the
majority of cases $p_c$ and $1/\mu_M$ differ by less than 50\%, but for the
remaining networks the discrepancy is larger, in some cases by more than one
order of magnitude.  These failures of prediction~(\ref{pc}) can be
understood by applying the knowledge acquired in the previous Sections.
Most (and the largest) of the violations occur when the NBC is localized on
small subgraphs, either overlapping hubs or the max K-core, which determine
the overall value of $\mu_M$.  In these cases the system actually undergoes
what can be seen as a double percolation
transition~\cite{Colomer-de-Simon2014}, reflected, in
Fig.~\ref{fig:susceptibility_left_peak}, by the presence of two distinct
peaks of the susceptibility $\chi_2(p)$ (see also
Ref.~\cite{HebertDufresne2019} for the effect of mesoscopic structures on
percolation).  In the networks considered in this figure, the
message-passing value $p=1/\mu_M$ signals the buildup of the connected
subgraph of relatively small size where NBC is localized, originating the
first susceptibility peak.  The second and largest peak occurs for much
larger values of $p$ and signals the formation of a percolating cluster
encompassing a larger fraction of the nodes.  Two (or even multiple) peaks
are present also in other networks.  The message-passing theory accurately
predicts only the leftmost of these peaks (see
Fig.~\ref{fig:susceptibility_left_peak}), while it does not give any
information about the position of other peaks and the associated transition.

Some other networks exhibit quite large discrepancies between $p_c$ and
$1/\mu_M$ but in the absence of a secondary peak.  Our theory does not
provide an explanation for these cases.  However, it must be remarked that
this phenomenology occurs for small networks, for which the very concept of
localization on a subgraph is not well defined.  Moreover, in these cases
the peak of the susceptibility is wide and it may hide the presence of
another peak (see Supplementary Figure SF3).

Finally, an ample discrepancy between $p_c$ and $1/\mu_M$ is observed also
for a few networks (\texttt{Road network TX}, \texttt{Road 512 network CA},
\texttt{Road network PA} and \texttt{US Power grid}) having very large
values of the average shortest path length $\av{\ell}$ and thus not
possessing the small-world property.  This is not surprising, as the almost
planar nature of these topologies makes our framework inapplicable to them.

In summary, realizing that localization of the NB centrality can
determine the value of $\mu_M$ for the whole structure allows us to
understand the presence of a double percolation transition in several
real-world networks. In these cases message-passing theory captures
only the first of the transitions, corresponding to the emergence
of a localized subgraph,
while the occurrence of the second transition is completely missed
by the theory\change{~\cite{Timar2017,Allard2019}.}

\section*{Discussion}
\label{sec:discussion}

Our results show that the non-backtracking centrality, which was
introduced to avoid the pathological self-reinforcement mechanism that
plagues standard eigenvector centrality, is affected by the same problem.
The NBC may also get localized on specific network subgraphs, with the same
bootstrap mechanism at work: Some nodes are highly central because they are
in ``contact'' with other central nodes and the latter are central because
they are in contact with the former.  The only difference is that for the
adjacency matrix the relevant subgraphs are stars and self-reinforcement
takes place among the hub and its direct neighbors~\cite{Castellano2017}.
For the NB matrix the
relevant subgraphs are groups of nodes sharing many neighbors and
self-reinforcement occurs at distance 2.  The possibility of localization
also for the NB matrix was overlooked so far, because it is exceedingly
unlikely in random uncorrelated networks.  However, as we show here, in
real-world topologies these structures are rather common. Indeed, cliques
and sets of overlapping hubs are, respectively, complete unipartite and
bipartite subgraphs, which naturally arise in many networks, for structural
or functional reasons. 

\change{The results presented here have a number of implications.  Which of
  the three contributions determines $\mu_M^\mathrm{global}$ in
  Eq.~(\ref{muprediction}) allows to rapidly estimate also the relevant
  non-backtracking centralities in the network. If $\mu_M^\mathrm{un}$
  dominates, then the NBC are given by Eq.~(\ref{eq:nbc_approximation}). If
  instead $\mu_M^\mathrm{oh}$ is largest, then non-backtracking centralities
  are given by Eq.~(\ref{nbcoh}) in the subset of overlapping hubs and are
  essentially zero elsewhere.  Similarly, when $\mu_M^\mathrm{core}$
  dominates in Eq.~(\ref{muprediction}), NBC is approximately constant in
  the max K-core and much smaller elsewhere. Additionally, our results allow
  to shed light of the LEV of the adjacency matrix, $\Lambda_M$.  In
  Ref.~\cite{Castellano2017}, it was argued that $\Lambda_M$ is determined
  by two subgraphs that have associated a large LEV, and that correspond to
  the node of maximum degree $k_\mathrm{max}$ (hub), taken as an isolated
  star graph, and the maximum $K$-core. Thus, in the spirit of Rayleigh's
  inequality~\cite{PVM_graphspectra}, it was proposed the approximation
  $\Lambda_M \simeq \max\{ \sqrt{\kmax}, \Lambda_{M}^\mathrm{core} \}$,
  where $\sqrt{\kmax}$ is the LEV of star graph of degree $\kmax$ and
  $\Lambda_{M}^\mathrm{core}$ is the LEV of the maximum $K$-core,
  approximated by its average degree
  $\av{k}_\mathrm{core}$~\cite{Castellano2017}. The subgraph composed by $n$
  overlapping hubs of degree $K$ turns out to possess also a large LEV of the
  adjacency matrix, given by $\Lambda_M^\mathrm{oh} = \sqrt{nK}$. We can
  then propose an improved approximation, taking into account the effect of
  overlapping hub, of the form   $\Lambda_M \simeq \max\{ \sqrt{\kmax},
  \Lambda_{M}^\mathrm{core}, \Lambda_M^\mathrm{oh}\}$. In Supplementary
  Figure~SF4 we check this new expression, observing
  that it provides some improvement in the estimation of the adjacency
  matrix LEV, particularly for networks of large size.}

The localization phenomenon of the NB matrix has \change{also} strong implications for
percolation and thus for the related susceptible-infected-removed model for
epidemic dynamics.  Quite surprisingly, this reveals strong analogies with
what happens in some regions of the phase-diagram of the paradigmatic
susceptible-infected-susceptible model for epidemic dynamics
(SIS)~\cite{Castellano2020}.  The formation (under appropriate conditions)
of localized clusters below the global epidemic transition is a striking
common feature of both types of dynamics, which they share despite their
completely different nature.  This intriguing similarity extends to the
predictive power of theoretical approaches. For SIS dynamics quenched
mean-field theory predicts when localized clusters of activity start to
appear, but misses the formation of an overall endemic
state~\cite{Castellano2020}.  For percolation (and SIR dynamics)
message-passing theory captures the formation of localized clusters but is
not predictive for what concerns the possible second transition involving a
much larger fraction of the network.  The quest for theoretical approaches
able to understand and predict this nontrivial second transition is a
challenging avenue for future research.

\change{Another related line for future research is the exploitation of the
  improved understanding presented here to devise targeted immunization
  strategies~\cite{Torres2020}.
  }
\section*{Methods}
\renewcommand{\thesubsection}{M\arabic{subsection}}

\subsection{Theory for uncorrelated networks}
\label{RadicchiMartin}

Denoting the PEV of the matrix $\mathbf{M}$ as $\vec{f} = \{ \vec{x},
\vec{w} \}$, we can rewrite Eq.~(\ref{eq:9}) as~\cite{PhysRevE.91.010801}
\begin{eqnarray}
  \sum_j A_{ij} x_j + w_i - k_i   w_i &=& \mu_M
  x_i, \\
  x_i &=& \mu_M w_i,
\end{eqnarray}
which translates into
\begin{equation}
  \mu_M \sum_j A_{ij} x_j + x_i - k_i   x_i = \mu_M^2
  x_i .
   \label{eq:basic}
\end{equation}
Summing over $i$ and rearranging, we obtain
\begin{equation}
  (\mu_M -1) \sum_i k_i x_i  =  (\mu_M^2 -1)  \sum_i x_i .
\end{equation}
Discarding the solution $\mu_M = 1$, which is always an eigenvalue, we have
\begin{equation}
  \sum_i k_i x_i  =  (\mu_M + 1)  \sum_i x_i ,
\end{equation}
leading to
\begin{equation}
  \mu_M = \frac{\sum_i k_i x_i}{\sum_i x_i} -1,
  \label{EqA6}
\end{equation}
which allows us to compute $\mu_M$ once the NBC is known.

Following Ref.~\cite{Martin2014}, we can obtain an approximation for the NB
matrix PEV (and hence for the NBC) by expanding the eigenvalue relation
\begin{equation}
  \mu_M v_{k \to l} = \sum_{i \to j}  B_{k \to l, i \to j} v_{i \to j},
  \label{eq:hashimoto}
\end{equation}
that, after some transformations can be written as~\cite{Martin2014}
\begin{equation}
  \mu_M v_{i \to l} = \sum_j A_{ij} (1 - \delta_{jl}) v_{j \to i} = \sum_{j
  \neq l} A_{ij}  v_{j \to i} . 
  \label{eq:eigenvalue_eq}
\end{equation}
Let us now compute the average value of $v_{i \to l}$ over all outgoing
nodes $i$ with a fixed degree $k_i = k$, that is 
\begin{equation}
  v_\mathrm{out}(k) =  \frac{1}{k N P(k)} \sum_{\substack{i \to l\\ k_i = k}}
  v_{i \to l} = \frac{1}{k N P(k)} \sum_{\substack{i,l \\ k_i = k}} A_{il}
  v_{i \to l},
  \label{eq:v_out}
\end{equation}
where $k N P(k)$ represents the number of edges emanating from nodes of
degree $k$.  Applying Eq.~\eqref{eq:eigenvalue_eq} to the previous equation
we can write
\begin{eqnarray}
  v_\mathrm{out}(k) &=& \frac{1}{k N P(k) \mu_M} \sum_{\substack{i,l\\ k_i =
  k}} \sum_{j \neq l} A_{ij} A_{il} v_{j \to i} \\ 
                    &=& \frac{1}{k N P(k)
  \mu_M} \sum_{\substack{i,j\\ k_i = k}} A_{ij} v_{j \to i} \sum_{l \neq j}
  A_{il} \\ 
                    &=& \frac{k - 1}{k N P(k) \mu_M} \sum_{\substack{i,j\\ k_i = k}}  
                    A_{ij} v_{j \to i}.
\end{eqnarray}
Assuming now~\cite{Martin2014} that the components $v_{j \to i}$ departing
from nodes of degree $k_i = k$ have the same distribution as in the whole
network (assumption valid in the limit of random uncorrelated networks), we
can substitute $v_{j \to i} \simeq \av{v} = \sum_{i \to j} v_{i \to j} /
(2E)$, where $E$ is the number of undirected edges in the original network.
With this assumption, we can write 
\begin{eqnarray}
  v_\mathrm{out}(k) &\simeq& \frac{\av{v}(k - 1)}{k N P(k) \mu_M}
  \sum_{\substack{i,j\\ k_i = k}}  A_{ij}  \nonumber \\
                    &=&  \frac{\av{v}(k - 1)}{k N P(k) \mu_M}
                    k N P(k) = \frac{\av{v}}{\mu_M} (k-1).
                    \label{eq:outgoing_average}
\end{eqnarray}

Analogously, we can compute the average of $v_{i \to l}$ over all ingoing
nodes $l$ with fixed degree $k_l = k$, 
\begin{equation}
  v_\mathrm{in}(k) =  \frac{1}{k N P(k)} \sum_{\substack{i \to l\\ k_l = k}}
  v_{i \to l} = \frac{1}{k N P(k)} \sum_{\substack{i,l \\ k_l = k}} A_{il}
  v_{i \to l}.
  \label{eq:v_in}
\end{equation}
Applying again Eq.~\eqref{eq:eigenvalue_eq}, we can write
\begin{eqnarray}
  v_\mathrm{in}(k) &=&  \frac{1}{k N P(k) \mu_M} \sum_{\substack{i, l\\ k_l = k}}
  \sum_{j \neq l} A_{il} A_{ij} v_{j \to i}  \nonumber\\
                   & \simeq&  \frac{\av{v}}{k N P(k) \mu_M}  \sum_{\substack{l\\ k_l = k}}
                 \sum_{j \neq l} \sum_i A_{li}A_{\change{ij}} \nonumber \\
                  & \simeq&  \frac{\av{v}}{k N P(k) \mu_M}  \sum_{\substack{l\\ k_l = k}}
                  \sum_{j \neq l}  \change{(A^2)}_{lj}.
\end{eqnarray}
The matrix element $\change{(A^2)}_{lj}$ counts the number of walks of length $2$
between nodes $l$ and $j$~\cite{Newman10}, and
$$
\change{
\sum_{\substack{l\\ k_l = k}} \sum_{j \neq l} (A^2)_{lj}
}
$$

counts those walks that start at nodes of
degree $k$ and are non-backtracking. In a tree-like network, the number of
such walks is equal to the number of next-nearest neighbors of nodes of
degree $k$, that is in average  $k N P(k) (\av{k^2} - \av{k}) /
\av{k}$~\cite{Newman10}. Therefore, we have
\begin{equation}
  v_\mathrm{in}(k) \simeq \frac{\av{v}}{\mu_M} \frac{\av{k^2} -
  \av{k}}{\av{k}}.
  \label{eq:ingoing_average}
\end{equation}

That is, in random uncorrelated networks, we have $v_\mathrm{out}(k) \sim k
- 1$ and $v_\mathrm{in}(k) \sim \mathrm{const.}$. Extending this relation at
the level of individual edges, we can approximate the normalized dependence
of the components of the NB matrix PEV as
\begin{equation}
  v_{i \to j} \simeq \frac{k_i - 1}{\sum_l k_l(k_l-1)}.
\end{equation}

In Supplementary Figure~SF5 we check
the dependence obtained for the components $v_{i \to j}$ of the PEV
of the NB matrix as a function of the outgoing $k_i$ and ingoing
$k_j$ degree, namely  $v_{i \to j} \sim k_i-1$.
The averaged components $v_\mathrm{out}$
and $v_\mathrm{in}$, defined in Eqs.~\eqref{eq:v_out} and~\eqref{eq:v_in},
correctly fulfill the scaling forms $v_\mathrm{out} \sim k -1$ and
$v_\mathrm{in} \sim \mathrm{const.}$, respectively. Indeed, for UCM
networks, the theoretical predictions in Eq.~\eqref{eq:outgoing_average}
and~Eq.~\eqref{eq:ingoing_average} are extremely well fulfilled.

\subsection{Localization of the non-backtracking centrality}
\label{appendix:localization}

The concept of vector localization/delocalization refers to whether the
components $x_i$ of a vector are evenly distributed over the network or they
attain a large value on some subset of nodes $V$ of size $N_V$ and are much
smaller in the rest of the network. In the first scenario we have $x_i \sim
\mathrm{const.}$ for all nodes $i$, and we say the vector is delocalized.
In the second scenario, one has $x_i \sim \mathrm{const.}$ for $i \in V$,
and  $x_i \sim  0$ for $i \notin V$, and we say the vector is localized on
$V$. For the NBC $x_i$, defined with a Euclidean normalization $\sum_i x_i^2
= 1$, localization can be measured in terms of the inverse participation
ratio $Y_4$~\cite{Goltsev2012,Martin2014}, defined as
\begin{equation}
  Y_4(N) = \sum_i x_i^4.
  \label{eq:ipr}
\end{equation}
For a delocalized vector, $x_i \sim N^{-1/2}$, so one has
$Y_4(N) \sim N^{-1}$; on the other hand, for a vector localized on a
subgraph of size
$N_V$, we have $Y_4(N) \sim N_V^{-1}$. Therefore, fitting the inverse
participation ratio to a power-law form $Y_4(N) \sim N^{-\alpha}$, a value
$\alpha \simeq 1$ indicates delocalization, while $\alpha < 1$ implies
localization on a subextensive set of nodes of size
$N_V \sim N^\alpha$~\cite{Pastor-Satorras2016}.
In the extreme case of localization on a finite set of nodes
(independent of $N$), one has instead $Y_4(N) \sim \mathrm{const.}$

The functional form derived for $x_i$ in Eq.~(\ref{eq:nbc_approximation})
helps to explain the localization properties of the NBC for UCM networks
observed in Ref.~\cite{Pastor-Satorras2016}.  In Supplementary
Figure~SF6 we show a comparison of the
inverse participation ratio $Y_4(N)$ numerically obtained in power-law UCM
networks with the theoretical prediction computed from
Eq.~\eqref{eq:nbc_approximation}, $Y_4^\mathrm{un}(N)$, and with the
prediction obtained from the annealed network approximation
Eq.~\eqref{eq:9}, $Y_4^\mathrm{an}(N)$. As we can see, the prediction from
our expression, $Y_4^\mathrm{un}(N)$, provides an almost perfect match for
the numerical observation, while the annealed network approximation exhibits
sizeable inaccuracies, particularly in the range $2.5 < \gamma < 3.5$.

\subsection{Largest non-backtracking eigenvalue of characteristic subgraphs}
\label{appendix:isolated}

\subsubsection*{Dangling star graph}

Let us consider a dangling star network, see Supplementary
Figure~SF1(a), formed by a hub $h$ of degree $K$ connected
to $K-1$ leaves $l$ of degree $1$ and by one edge to a connector node $n$
of a generic network.  By applying Eq.~\eqref{eq:basic}, we obtain the
following equations for the LEV $\mu_M$ and the NBC: 
\begin{eqnarray}
  \mu_M [ (K-1)x_l + x_n ] - (K-1)x_h &=& \mu_M^2 x_h,\\
  \mu_M x_h &=&   \mu_M^2 x_l,\\
  \mu_M [\sum_{\change{i \ne h}} A_{ni} x_i + x_h] - k_n x_n  &=&   \mu_M^2 x_n,
\end{eqnarray}
where $k_n$ is the degree of node $n$, $x_l$ is the NBC centrality of
each leaf, and the equations corresponding to the rest of the nodes
$i \neq n$ are the same as in the absence of the dangling star.  

From the first two equations, assuming $\mu_M \neq 0$, we obtain $x_h =
\mu_M x_l$ and $x_n = \mu_M x_h$.  Introducing the last equality into the
third equation, the dependence on $x_h$ drops out and the equation takes the
form of Eq.~\eqref{eq:basic} in the absence of the dangling star.  We
conclude therefore that a dangling star is unable to alter the value of the
overall LEV $\mu_M$ and its NBC depends only on the centrality of the
connector node $n$.  The reason for this is the absence of non-backtracking
paths between the hub and the leaves, so that the hub has the effect of a
node of degree one~\cite{Martin2014,Krzakala2013}.

\subsubsection*{Integrated star graph}

The case of an integrated star of degree $K$, i.e., a star connected by $K$
edges to $K$ randomly chosen connector nodes in a network,
Supplementary Figure~SF1(b), is more difficult to analyze.  To simplify
calculations, we consider the case of a regular network with fixed degree
$q$. For symmetry reasons, the nodes connected to the hub, of degree $q+1$,
have approximately the same NBC, $x_1$, different from the centrality $x_2$
of the nodes not connected to the hub, and also from $x_0$, the centrality
of the hub.  Applying the Ihara-Bass determinant formula,
Eq.~\eqref{eq:basic}, we can write  
\begin{eqnarray*}
  \mu_M &&K x_1 = (K + \mu_M^2 - 1) x_0,\\
  \mu_M &&\left[x_0 + q\frac{K}{N}  x_1 + q\left(1-\frac{K}{N}
  \right)x_2 \right]= 
  (q  + \mu_M^2) x_1,\\
  \mu_M &&\left[q\frac{K}{N}  x_1 + q \left(1-\frac{K}{N} \right)
  x_2\right]=
  (q  + \mu_M^2 -1) x_2,\\
\end{eqnarray*}
where\change{, to ease calculations,} we have made the mean-field assumption that nodes in the network are
neighbors of nodes connected to the hub with probability $K/N$, and
otherwise with probability $1 - K/N$\change{, which is valid in the limit of
large $K$ and $N$}. These conditions lead to the equation
for $\mu_M$

\begin{equation}
  \left.
    \begin{aligned}
      \mu_M^{5} + \mu_M^{4} \left(1 - q\right) & + \mu_M^{3} \left(q - 1\right)  
  - \mu_M^{2}  \left[ \frac{Kq(K-1)}{N} + (q-1)^2 \right]\\
                                               &+ \mu_M \frac{q(K-1)(N-K)}{N} - q(K-1)(q-1) = 0
  \end{aligned} \right. ,
\label{eq36}
\end{equation}
where we have factorized the trivial solution $\mu_M = 1$. 
This is an algebraic equation of fifth order than cannot be solved
analytically in general.
However, for $K(K-1)q \gg N$, assuming $\mu_M \gg q-1$, it reduces to
\begin{equation}
  \mu_M^5 +  \mu_M^2 \frac{Kq(K-1)}{N}= 0,
\end{equation}
leading to the solution
\begin{equation}
  \mu_M^\mathrm{h} \simeq \left(\frac{q K(K-1)}{N} \right)^{1/3}.
  \label{eq:integratedhub}
\end{equation}
Instead for $K(K-1)q \ll N$, assuming $\mu_M = q-1+\epsilon$
and expanding Eq.~(\ref{eq36}) to first order in $\epsilon$, we obtain
\be
\epsilon = \frac{(q-1)^2+(q-1)}{(q-1)^4+(q-1)^3+q(K-1)} \frac{Kq(K-1)}{N}.
\ee
Hence the value of $\mu_M$ is very close to the value $q-1$ of
the original random regular network, with a correction that
vanishes with $N$.
We conclude that the addition of a finite
integrated hub does not change the value $\mu_M$ of the whole network
unless $K(K-1)q \gg N$, a case which may be relevant in small networks.
Not surprisingly, the uncorrelated expression
Eq.~(\ref{eq:lev_approximation}) fails here, since it predicts a finite
value $\mu_M^\mathrm{un} \sim 2 q$, in the limit of large $K$.  

While we considered a star integrated into a homogeneous network,
Supplementary Figure~SF7 shows that the same picture
is valid also in the case of power-law distributed synthetic networks,
replacing $q$ by the network average degree $\av{k}$: for $K$ up to
values of the order of $(N/\av{k})^{1/2}$ the addition of the hub
has no effect on $\mu_M$; for larger values, Eq.~(\ref{eq:integratedhub})
holds.

\subsubsection*{Overlapping hubs}

Let us consider now a graph composed of $n$ hubs, sharing all their $K$
leaves, see Supplementary Figure~SF1(c).  We can evaluate
$\mu_M$ and $x_i$ by applying again the Ihara-Bass determinant formula. For
symmetry reasons, the components $x_h$ of the hubs are equal, and
correspondingly the components $x_\ell$ of the leaves. Thus, from
Eq.~\eqref{eq:basic} we can write
\begin{eqnarray}
  \mu_M K x_\ell &=& (K + \mu_M^2 - 1) x_h,\\
  \mu_M n x_h &=& (n  + \mu_M^2 - 1) x_\ell,
\end{eqnarray}
Imposing that the components $x_h$  and $x_\ell$ are non-zero, we obtain the
largest eigenvalue
\begin{equation}
  \mu_M^\mathrm{oh}  = \sqrt{(n-1)(K-1)},
\end{equation}
while the NB centralities fulfill
\begin{equation}
  \frac{x_\ell^2}{x_h^2} = \frac{K-1}{K^2} \frac{n^2}{n-1}.
  \label{nbcoh}
\end{equation}
That is, for large $K$, the NBC becomes strongly localized in the hubs.

In Supplementary Figure~SF8 we check the effects of
adding $n$ overlapping hubs of degree $\change{K}$ to power-law distributed synthetic
networks.  As we can see, as soon as $\mu_M^\mathrm{oh}$ is large enough (in
practice, when $K > 1 + \left( \frac{\av{k^2}}{\av{k}} -1 \right
  )^2/(n-1)$), the actual value of the NB LEV is dominated by the presence of
  the overlapping hubs.

\subsection{$K$-core decomposition}
\label{appendix:kcores}

The $K$-core decomposition~\cite{Seidman1983269} is an iterative
classification process of the vertices of a network in layers of increasing
density of mutual connections, 
\change{denoted by increasing values of the index $K$}.
One starts removing  the vertices of degree
$k=1$, repeating the process until only nodes with degree $k  \geq 2$ are
left. The removed nodes constitute the $K = 1$ shell, and the remaining ones
are the $K = 2$ core.
At the next step, all vertices with degree $k=2$ are
iteratively removed, thus leaving the $K = 3$ core. The procedure is
repeated  until the  maximum $K$-core (of index $K_M$) is reached, such that
one more iteration removes all nodes in the network. The maximum $K$-core of
generic networks is usually a homogeneous subgraph~\cite{Castellano2017}.
The $K$-core structure of networks has been proposed as a classification of
node importance in dynamical processes on complex topologies~\cite{kitsak2010}.

\subsection{Algorithm to determine optimal $n$ and $K$ values for overlapping
hubs}
\label{appendix:heuristicalgorithm}

The determination of the set of all overlapping hubs in a real-world network
is highly time consuming.  We can however obtain a working approximation
using the following greedy algorithm: We order the nodes in decreasing order
of their degree, $i_1, i_2, \ldots, i_N$. Starting from node $i_\alpha$, we
visit the set of nodes
\change{ $i_\alpha, i_{\alpha+1}, \ldots i_{\alpha+q}$ and identify
and identify the number of common neighbors $K_q^\alpha$,}
that are common neighbors of the set of nodes $i_\alpha, i_{\alpha+1}, \ldots i_{\alpha+q}$.
Repeating this process for all nodes in the network, we compute the values
$K_q^\alpha$ for all nodes $\alpha$ and all sets of  nodes (in decreasing
order of degree) of length $q +1$. We choose as values of $n$ and $K$ the
values of $q+1$ and $K_q^\alpha$ that maximize the product
$q (K_q^\alpha-1)$.

\subsection{Numerical simulations of bond percolation}
\label{perc}

We consider the bond percolation process in which network edges are
randomly kept with probability $p$ and removed with probability $1-p$. For
each realization of this process with a given value of $p$, one considers the
largest cluster remaining in the network, of size $S_p$. The average of this
quantity over independent realization is denoted by $\av{S_p}$. The
critical percolation point $p_c$ separates a subcritical phase at $p < p_c$,
in which only clusters of small size are present, so that $\av{S_p} / N \to
0$ in the thermodynamic limit $N\to\infty$, from a supercritical phase at $p
> p_c$, in which there is a finite spanning cluster leading to
$\av{S_p} /N \to \mathrm{const.}$~\cite{stauffer94}. 

In order to estimate the value of the percolation point, one considers the
susceptibility $\chi_2(p)$, defined
as~\cite{Castellano2016,PhysRevE.91.010801}
\begin{equation}
  \chi_2(p) = \frac{\av{S_p^2} - \av{S_p}^2}{\av{S_p}}.
\end{equation}
The percolation threshold $p_c$ is defined as the value of $p$ for which
$\chi_2(p)$ shows a maximum~\cite{Castellano2016}.  To compute numerically
$\chi_2(p)$ in real-world networks we perform the averages on bond
percolation experiments applying the Newman-Ziff
algorithm~\cite{Newman2000}.

\section*{Acknowledgments} C. C. thanks Abolfazl Ramezanpour for useful
comments and suggestions.  We acknowledge financial support from the Spanish
Government's MINECO, under project FIS2016-76830-C2-1-P \change{and MICINN,
under project PID2019-106290GB-C21.}
  
\section*{Author contributions}
Both authors designed the research and developed the theoretical analysis.
R. P.-S. performed the numerical analysis.  Both authors analyzed the
results and wrote the paper.

\section*{Competing interests}
The authors declare no competing interests.

\bibliographystyle{mynature}


\clearpage

\begin{center}
\bf \Large Supplementary Information
\end{center}

\renewcommand{\thetable}{\textbf{ST\arabic{table}}}
\renewcommand{\thefigure}{\textbf{SF\arabic{figure}}}

\setcounter{table}{0}
\setcounter{figure}{0}
\setcounter{section}{0}

\renewcommand{\figurename}{\textbf{Supplementary Figure}}
\renewcommand{\tablename}{\textbf{Supplementary Table}}

\clearpage

\section*{Supplementary Table}


\begin{center}
  \renewcommand*{\arraystretch}{1.3}
  \setlength{\LTcapwidth}{0.9\linewidth}
  \begin{longtable}{@{\extracolsep{\fill}}ll|ccc|ccccc|c@{}}
    \caption{Topological and spectral properties of the $109$ real-world
      networks in Ref.~[18] for which we test our theory.  We
      report the following properties of this set of networks: $N$: network
      size; $\av{k}$: average degree; $k_\mathrm{max}$: maximum degree;
      $\mu_M$: LEV of the NBC;  $\mu_M^\mathrm{an}$: theoretical
      approximation for  $\mu_M$ within the annealed network approximation,
      Eq.~(9); $\mu_M^\mathrm{un}$: theoretical
      approximation for  $\mu_M$ in uncorrelated networks,
      Eq.~(8); $\mu_M^\mathrm{oh}$: theoretical
      approximation for $\mu_M$ taking into account the effect of
      overlapping hubs, Eq.~(10); $\mu_M^\mathrm{core}$: LEV of
      the NBC for the maximum $K$-core of the network;  $p_c^{-1}$: inverse
      of the numerical percolation threshold $p_c$, estimated as the
    position of the principal peak of the susceptibility $\chi_2$.}
    \label{netdetails}
    \\
\hline\hline
 & Network                  &     $N$ &   $\av{k}$ &   $k_\mathrm{max}$ &   $\mu_M$ &   $\mu_M^\mathrm{an}$ &   $\mu_M^\mathrm{un}$ &   $\mu_M^\mathrm{oh}$ &   $\mu_M^\mathrm{core}$ &   $p_c^{-1}$ \\
\hline
\endfirsthead
\hline\hline
 & Network                  &     $N$ &   $\av{k}$ &   $k_\mathrm{max}$ &   $\mu_M$ &   $\mu_M^\mathrm{an}$ &   $\mu_M^\mathrm{un}$ &   $\mu_M^\mathrm{oh}$ &   $\mu_M^\mathrm{core}$ &   $p_c^{-1}$ \\
\hline
\endhead

	     0 & Social 3                 &      32 &   5.00 &     13 &    4.74 &     4.94 &   4.76 &   1.41 &   3.96 &    4.0572 \\
	     1 & Karate club              &      34 &   4.59 &     17 &    5.29 &     6.77 &   4.75 &   2.00 &   4.18 &    4.3472 \\
	     2 & Protein 2                &      53 &   4.64 &      8 &    4.68 &     4.39 &   4.52 &   1.73 &   4.29 &    2.9545 \\
	     3 & Dolphins                 &      62 &   5.13 &     12 &    5.99 &     5.81 &   5.75 &   2.00 &   5.74 &    4.1694 \\
	     4 & Social 1                 &      67 &   4.24 &     11 &    4.36 &     4.25 &   4.38 &   1.73 &   3.00 &    3.2801 \\
	     5 & Les Miserables           &      77 &   6.60 &     36 &   10.75 &    11.06 &  10.04 &   5.29 &   9.40 &    7.5977 \\
	     6 & Protein 1                &      95 &   4.48 &      7 &    4.25 &     3.95 &   4.01 &   1.41 &   4.23 &    1.7169 \\
	     7 & E. Coli, transcription   &      97 &   4.37 &     10 &    5.34 &     4.41 &   4.86 &   1.73 &   4.83 &    1.9855 \\
	     8 & Political books          &     105 &   8.40 &     25 &   10.63 &    10.93 &  10.40 &   3.61 &   8.97 &    5.4306 \\
	     9 & David Copperfield        &     112 &   7.59 &     49 &   11.54 &    12.77 &  11.44 &   4.47 &  10.32 &    9.2696 \\
	    10 & College football         &     115 &  10.66 &     12 &    9.77 &     9.73 &   9.75 &   3.46 &   9.75 &    7.3987 \\
	    11 & S 208                    &     122 &   3.10 &     10 &    2.75 &     2.77 &   2.76 &   1.00 &   2.75 &    2.1438 \\
	    12 & High school, 2011        &     126 &  27.13 &     55 &   32.85 &    31.79 &  32.13 &   9.59 &  26.09 &   25.8514 \\
	    13 & Bay Dry                  &     128 &  32.42 &    110 &   38.44 &    39.11 &  38.04 &  13.86 &  34.69 &   31.7995 \\
	    14 & Bay Wet                  &     128 &  32.91 &    110 &   38.91 &    39.50 &  38.53 &  12.37 &  33.64 &   32.3371 \\
	    15 & Radoslaw Email           &     167 &  38.92 &    139 &   59.43 &    63.46 &  58.35 &  26.72 &  52.81 &   49.4981 \\
	    16 & High school, 2012        &     180 &  24.67 &     56 &   29.01 &    28.55 &  28.72 &   6.93 &  22.69 &   22.5561 \\
	    17 & Little Rock Lake         &     183 &  26.60 &    105 &   40.06 &    41.89 &  38.37 &  16.12 &  34.70 &   32.4229 \\
	    18 & Jazz                     &     198 &  27.70 &    100 &   38.82 &    37.64 &  37.83 &  10.39 &  28.00 &   30.6470 \\
	    19 & S 420                    &     252 &   3.17 &     14 &    2.89 &     2.91 &   2.90 &   1.00 &   2.89 &    2.2160 \\
	    20 & C. Elegans, neural       &     297 &  14.46 &    134 &   22.76 &    25.05 &  20.87 &   7.35 &  20.02 &   18.1451 \\
	    21 & Network Science          &     379 &   4.82 &     34 &    8.71 &     7.02 &   6.53 &   4.00 &   7.00 &    2.5558 \\
	    22 & Dublin                   &     410 &  13.49 &     50 &   22.24 &    17.72 &  18.75 &   6.00 &  21.31 &   12.9029 \\
	    23 & US Air Trasportation     &     500 &  11.92 &    145 &   46.54 &    52.78 &  43.04 &  15.30 &  31.58 &   37.8733 \\
	    24 & S 838                    &     512 &   3.20 &     22 &    2.94 &     3.03 &   2.96 &   1.00 &   2.94 &    2.2063 \\
	    25 & Yeast, transcription     &     662 &   3.21 &     71 &    6.50 &    12.51 &   3.02 &   5.57 &   5.09 &    4.1265 \\
	    26 & URV email                &    1133 &   9.62 &     71 &   19.27 &    17.69 &  18.37 &   4.00 &  10.00 &   15.4127 \\
	    27 & Political blogs          &    1222 &  27.36 &    351 &   72.56 &    80.26 &  66.72 &  13.82 &  42.71 &   58.9779 \\
	    28 & Air traffic              &    1226 &   3.93 &     34 &    7.48 &     6.36 &   6.26 &   2.24 &   6.70 &    5.4898 \\
	    29 & Yeast, protein           &    1458 &   2.67 &     56 &    5.05 &     6.13 &   3.25 &   1.00 &   4.00 &    3.3193 \\
	    30 & Petster, hamster         &    1788 &  13.96 &    272 &   44.31 &    44.55 &  40.19 &  14.00 &  34.73 &   36.3558 \\
	    31 & UC Irvine                &    1893 &  14.62 &    255 &   46.25 &    54.64 &  43.70 &   9.27 &  35.39 &   39.2625 \\
	    32 & Yeast, protein           &    2172 &   6.05 &    215 &   18.54 &    18.79 &  16.31 &   4.36 &  10.66 &   13.6990 \\
	    33 & Japanese                 &    2698 &   5.93 &    725 &   38.16 &   107.61 &  23.46 &  14.32 &  23.56 &   30.7833 \\
	    34 & Open flights             &    2905 &  10.77 &    242 &   61.33 &    54.84 &  57.28 &  14.14 &  32.02 &   49.2010 \\
	    35 & GR-QC, 1993-2003         &    4158 &   6.46 &     81 &   44.44 &    16.98 &  27.64 &  20.20 &  42.00 &    7.4308 \\
	    36 & Tennis                   &    4338 &  37.74 &    451 &  160.17 &   157.91 & 158.09 &  17.38 & 124.14 &  136.0226 \\
	    37 & US Power grid            &    4941 &   2.67 &     19 &    6.23 &     2.87 &   2.88 &   1.41 &   5.06 &    1.5142 \\
	    38 & HT09                     &    5352 &   6.91 &   1287 &   41.01 &   198.98 &   9.06 &  13.27 &  25.42 &   34.8533 \\
	    39 & Hep-Th, 1995-1999        &    5835 &   4.74 &     50 &   17.01 &     8.12 &   9.41 &   6.48 &  17.00 &    9.0419 \\
	    40 & Reactome                 &    5973 &  48.81 &    855 &  206.88 &   142.31 & 160.58 &  91.04 & 197.41 &   87.9832 \\
	    41 & Jung                     &    6120 &  16.43 &   5655 &  128.35 &   990.77 &  29.33 & 103.36 &  77.46 &  107.0054 \\
	    42 & Gnutella, Aug.  8, 2002  &    6299 &   6.60 &     97 &   26.51 &    16.66 &  17.60 &   4.69 &  22.35 &   22.0829 \\
	    43 & JDK                      &    6434 &  16.68 &   5923 &  129.28 &   981.71 &  29.92 & 103.74 &  77.46 &  107.1269 \\
	    44 & AS Oregon                &    6474 &   3.88 &   1458 &   35.04 &   163.81 &  14.68 &  18.52 &  14.96 &   28.0308 \\
	    45 & English                  &    7377 &  11.98 &   2568 &  104.34 &   319.70 &  59.17 &  32.83 &  58.34 &   87.9832 \\
	    46 & Gnutella, Aug.  9, 2002  &    8104 &   6.42 &    102 &   26.56 &    15.82 &  16.65 &   5.10 &  23.39 &   21.9957 \\
	    47 & French                   &    8308 &   5.74 &   1891 &   52.46 &   217.01 &  26.58 &  18.49 &  23.12 &   43.1321 \\
	    48 & Hep-Th, 1993-2003        &    8638 &   5.74 &     65 &   30.01 &    11.99 &  14.42 &  13.75 &  30.00 &   13.2956 \\
	    49 & Gnutella, Aug.  6, 2002  &    8717 &   7.23 &    115 &   20.47 &    13.40 &  14.02 &   8.37 &  16.94 &   15.0067 \\
	    50 & Gnutella, Aug.  5, 2002  &    8842 &   7.20 &     88 &   21.58 &    13.79 &  14.01 &   5.29 &  18.62 &   17.2084 \\
	    51 & PGP                      &   10680 &   4.55 &    205 &   41.03 &    17.88 &  26.19 &   9.49 &  35.73 &   14.6018 \\
	    52 & Gnutella, August 4 2002  &   10876 &   7.35 &    103 &   15.28 &    12.97 &  12.86 &   4.36 &  13.19 &   12.9654 \\
	    53 & Hep-Ph, 1993-2003        &   11204 &  21.00 &    491 &  243.75 &   129.88 & 206.61 & 113.67 & 237.00 &  209.4101 \\
	    54 & Spanish                  &   11558 &   7.45 &   2986 &   93.51 &   456.58 &  40.68 &  32.19 &  44.11 &   78.1537 \\
	    55 & DBLP, citations          &   12495 &   7.93 &    709 &   38.06 &    42.77 &  33.58 &  14.56 &  31.19 &   30.2164 \\
	    56 & Spanish                  &   12643 &   8.70 &   5169 &  100.13 &   806.66 &  28.19 &  35.37 &  47.63 &   83.9067 \\
	    57 & Cond-Mat, 1995-1999      &   13861 &   6.44 &    107 &   23.14 &    12.54 &  14.83 &   6.00 &  16.00 &   15.5067 \\
	    58 & Astrophysics             &   14845 &  16.12 &    360 &   72.21 &    44.46 &  55.60 &  19.60 &  55.00 &   55.0700 \\
	    59 & Google                   &   15763 &  18.85 &  11401 &  156.61 &   900.63 &  47.71 &  86.99 & 106.57 &  125.5953 \\
	    60 & AstroPhys, 1993-2003     &   17903 &  22.00 &    504 &   92.54 &    64.70 &  77.74 &  16.55 &  55.00 &   76.5451 \\
	    61 & Cond-Mat, 1993-2003      &   21363 &   8.55 &    279 &   35.80 &    21.47 &  26.02 &  12.73 &  24.00 &   27.3670 \\
	    62 & Gnutella, Aug.  25, 2002 &   22663 &   4.83 &     66 &    9.38 &     9.75 &   8.96 &   2.45 &   8.87 &    8.6900 \\
	    63 & Internet                 &   22963 &   4.22 &   2390 &   64.68 &   260.46 &  28.28 &  24.25 &  39.97 &   51.4491 \\
	    64 & Thesaurus                &   23132 &  25.69 &   1062 &   97.70 &   102.29 &  94.53 &  15.17 &  82.91 &   88.5512 \\
	    65 & Cora                     &   23166 &   7.70 &    377 &   29.28 &    22.68 &  19.42 &   8.06 &  16.91 &   21.9551 \\
	    66 & Linux, mailing list      &   24567 &  12.88 &   2989 &  220.15 &   339.98 & 178.45 &  46.66 & 121.12 &  190.7713 \\
	    67 & AS Caida                 &   26475 &   4.03 &   2628 &   59.41 &   279.24 &  26.29 &  24.62 &  34.41 &   48.1394 \\
	    68 & Gnutella, Aug.  24, 2002 &   26498 &   4.93 &    355 &   10.78 &    11.03 &  10.77 &   2.24 &  10.34 &    9.4122 \\
	    69 & Hep-Th, citations        &   27400 &  25.69 &   2468 &  106.82 &   105.40 &  88.46 &  54.94 &  43.36 &   90.5805 \\
	    70 & Cond-Mat, 1995-2003      &   27519 &   8.44 &    202 &   38.30 &    21.29 &  26.52 &  12.45 &  23.00 &   29.3631 \\
	    71 & Digg                     &   29652 &   5.72 &    283 &   27.63 &    27.07 &  27.22 &   4.24 &  23.52 &   24.1307 \\
	    72 & Linux, soft.             &   30817 &  13.84 &   9338 &  154.98 &   851.62 &  34.55 &  69.53 &  58.94 &  129.6788 \\
	    73 & Enron                    &   33696 &  10.73 &   1383 &  115.48 &   141.36 &  90.59 &  15.30 &  79.03 &   99.2556 \\
	    74 & Hep-Ph, citations        &   34401 &  24.46 &    846 &   74.33 &    62.50 &  61.91 &  19.85 &  33.57 &   63.3955 \\
	    75 & Cond-Mat, 1995-2005      &   36458 &   9.42 &    278 &   49.17 &    26.88 &  34.58 &  14.66 &  28.00 &   38.8247 \\
	    76 & Gnutella, Aug.  30, 2002 &   36646 &   4.82 &     55 &   11.39 &    10.46 &   9.93 &   2.83 &   6.00 &   10.2897 \\
	    77 & Slashdot                 &   51083 &   4.56 &   2915 &   44.95 &    80.57 &  34.72 &   9.49 &  35.63 &   37.7581 \\
	    78 & Gnutella, Aug.  31, 2002 &   62561 &   4.73 &     95 &   11.48 &    10.60 &  10.05 &   2.00 &   9.57 &   10.4354 \\
	    79 & Facebook                 &   63392 &  25.77 &   1098 &  130.82 &    87.05 & 105.41 &  12.37 & 100.56 &  114.5491 \\
	    80 & Epinions                 &   75877 &  10.69 &   3044 &  181.65 &   182.88 & 161.62 &  23.22 & 129.54 &  161.1385 \\
	    81 & Slashdot zoo             &   79116 &  11.82 &   2534 &  127.57 &   145.30 & 106.05 &  23.94 &  80.90 &  112.1438 \\
	    82 & Flickr                   &  105722 &  43.83 &   5425 &  614.42 &   348.21 & 429.50 &  68.08 & 572.00 &   71.8799 \\
	    83 & Wikipedia, edits         &  113123 &  35.82 &  20153 &  389.69 &   688.54 & 289.34 &  96.31 & 216.89 &  347.9713 \\
	    84 & Petster, cats            &  148826 &  73.21 &  80634 & 1160.43 &  9291.62 & 261.40 & 873.92 & 664.31 & 1017.3354 \\
	    85 & Gowalla                  &  196591 &   9.67 &  14730 &  159.86 &   305.58 &  76.47 &  35.33 &  81.48 &  136.8895 \\
	    86 & Libimseti                &  220970 & 155.98 &  33389 &  943.38 &  1639.96 & 671.28 & 140.18 & 572.24 &  882.0520 \\
	    87 & EU email                 &  224832 &   3.02 &   7636 &   97.09 &   566.65 &  26.93 &   9.85 &  72.95 &   82.2030 \\
	    88 & Web Stanford             &  255265 &  15.21 &  38625 &  423.82 &  2029.74 &  46.88 & 336.93 & 130.43 &   18.0571 \\
	    89 & Amazon, Mar.  2, 2003    &  262111 &   6.87 &    420 &   17.80 &    10.14 &  10.04 &   5.10 &   7.92 &   10.5122 \\
	    90 & DBLP, collaborations     &  317080 &   6.62 &    343 &  114.72 &    20.75 &  31.33 &  40.00 & 112.00 &   29.5135 \\
	    91 & Web Notre Dame           &  325729 &   6.69 &  10721 &  175.66 &   279.68 &  55.53 &  70.99 & 164.69 &   12.3073 \\
	    92 & MathSciNet               &  332689 &   4.93 &    496 &   33.53 &    15.43 &  18.85 &   6.71 &  23.00 &   21.0558 \\
	    93 & CiteSeer                 &  365154 &   9.43 &   1739 &   52.54 &    47.45 &  29.49 &  10.25 &  35.33 &   40.4606 \\
	    94 & Zhishi                   &  372840 &  12.43 & 127066 &  942.98 & 27908.59 &  15.43 & 942.62 & 295.29 &   33.1929 \\
	    95 & Actor coll.  net.        &  374511 &  80.18 &   3956 &  847.55 &   417.32 & 573.14 &  61.48 & 592.15 &  776.8318 \\
	    96 & Amazon, Mar.  12, 2003   &  400727 &  11.73 &   2747 &   35.03 &    29.33 &  20.30 &  16.25 &  31.68 &   24.9309 \\
	    97 & Amazon, Jun.  6, 2003    &  403364 &  12.11 &   2752 &   40.31 &    29.55 &  21.73 &  17.15 &  33.06 &   27.5921 \\
	    98 & Amazon, May 5, 2003      &  410236 &  11.89 &   2760 &   40.36 &    29.93 &  21.81 &  17.38 &  32.59 &   27.6319 \\
	    99 & Petster, dogs            &  426485 &  40.06 &  46503 &  734.01 &  2054.76 & 363.83 & 427.47 & 427.52 &  665.8499 \\
	   100 & Road network PA          & 1087562 &   2.83 &      9 &    3.11 &     2.20 &   2.24 &   1.41 &   2.90 &    1.4442 \\
	   101 & YouTube friend.  net.    & 1134890 &   5.27 &  28754 &  185.14 &   493.53 &  80.41 &  56.99 & 105.78 &  156.6161 \\
	   102 & Road network TX          & 1351137 &   2.78 &     12 &    3.56 &     2.15 &   2.19 &   1.41 &   3.51 &    1.3623 \\
	   103 & AS Skitter               & 1694616 &  13.09 &  35455 &  653.66 &  1444.15 &  89.41 & 260.77 & 154.76 &  563.5708 \\
	   104 & Road network CA          & 1957027 &   2.82 &     12 &    3.32 &     2.17 &   2.21 &   1.41 &   3.17 &    1.4409 \\
	   105 & Wikipedia, pages         & 2070367 &  40.90 & 230040 &  775.44 &  3345.71 & 308.90 & 190.21 & 302.35 &  699.8880 \\
	   106 & US Patents               & 3764117 &   8.77 &    793 &  110.45 &    20.34 &  27.28 &  55.41 &  75.82 &   34.3938 \\
	   107 & DBpedia                  & 3915921 &   6.42 & 469692 &  462.92 & 13856.37 &  17.53 & 388.33 &  28.30 &   58.5652 \\
	   108 & LiveJournal              & 5189808 &  18.76 &  15016 &  537.93 &   154.42 & 221.84 &  43.89 & 408.16 &  361.7945 \\

\hline\hline

  \end{longtable}
\end{center}

\clearpage

\clearpage

%
%

\clearpage

\section*{Supplementary Figures}

\begin{figure}[h]
  \centering \includegraphics[width=0.7\columnwidth]{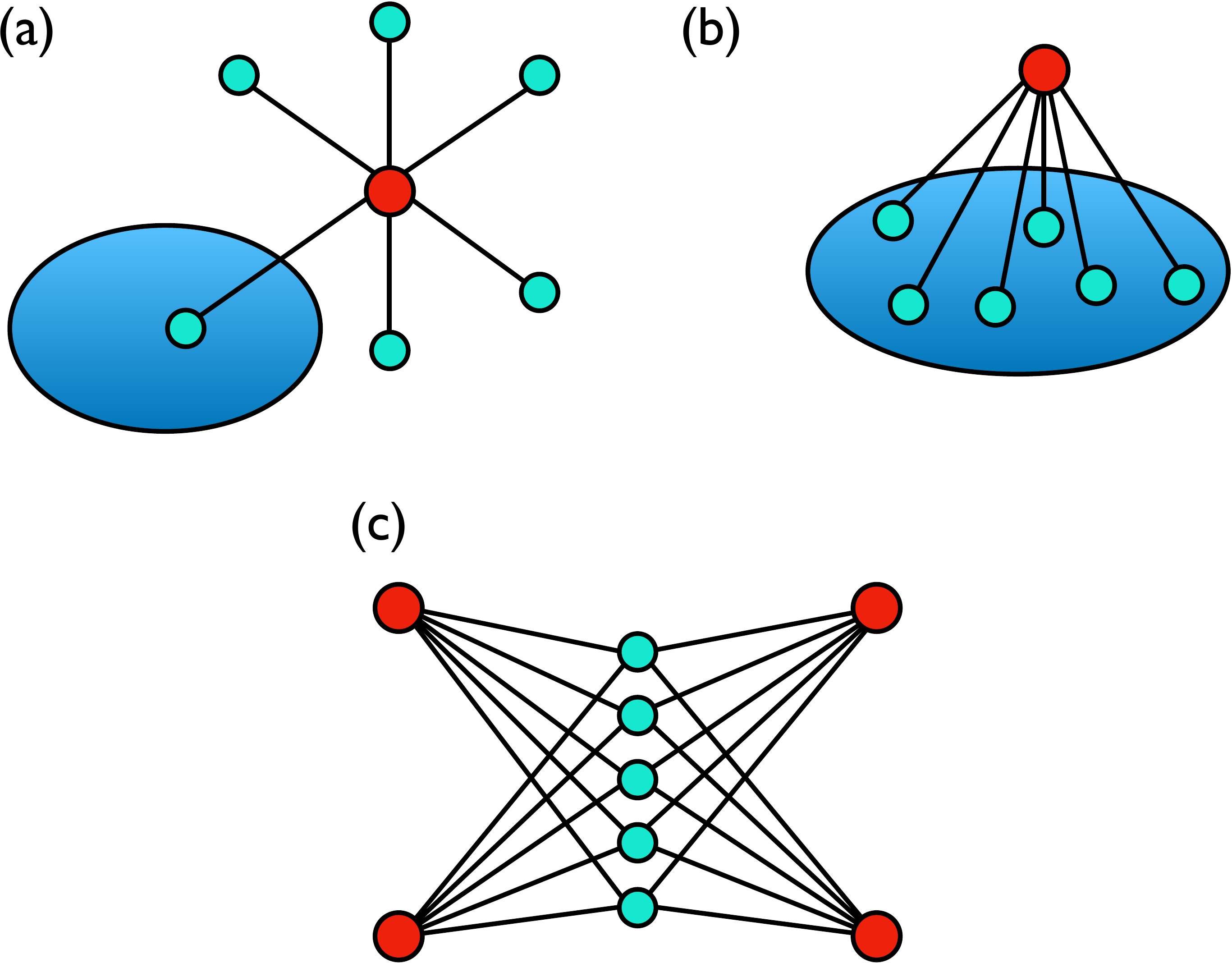}
  \caption{{\bf Graphical representation of star subgraphs.} (a) Dangling
    hub of degree $K$, connected to $K-1$ leaves of degree $1$ and to a
    connector node in a generic network. $K=6$. (b) Integrated hub of degree
  $K$ connected to $K$ connector nodes in a generic network. $K=6$. (c)
Example of $n$ overlapping hubs of degree $K$, sharing the same set of
leaves of degree $n$. $n=4$, $K=5$.}
  \label{fig:stargraphs}
\end{figure}

\begin{figure}[h]
  \centering
  \includegraphics[width=0.7\columnwidth]{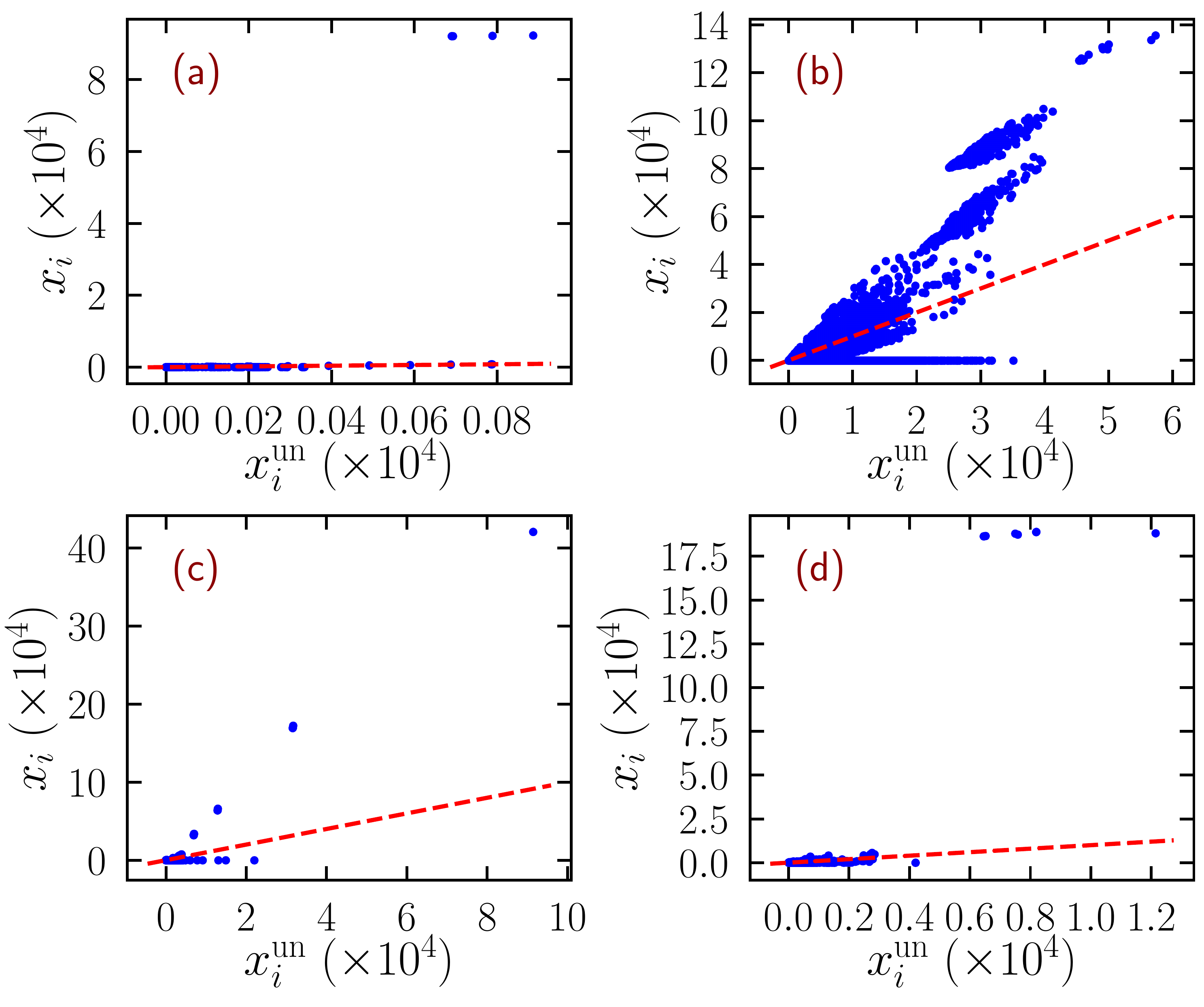}
  \caption{{\bf NBC localization in real-world networks.} Scatter plot of
    the NBC $x_i$ as a function of the theoretical prediction
    $x_i^\mathrm{un}$, Eq.~(7) in four examples of
  real-world networks (a) \texttt{Zhishi}; (b) \texttt{Flickr}; (c)
\texttt{Web Notre Dame}; (d) \texttt{Web Stanford}.  The dashed line
represents the behavior $y=x$.}
  \label{fig:real_failures_examples}
\end{figure}
\begin{figure}[h]
  \centering
  \includegraphics[width=0.88\columnwidth]{All_Susceptibilities_Chi2.pdf}
  \caption{{\bf Susceptibility $\chi_2(p)$ for all 109 networks considered.}
    In each plot, the green dashed vertical line(s) denote the position(s)
    of the peak(s), the black continuous vertical line denotes the value of
  $1/\mu_M$.}
  \label{GuinnessFigure}
\end{figure}

\begin{figure}
  \centering \includegraphics[width=0.9\columnwidth]{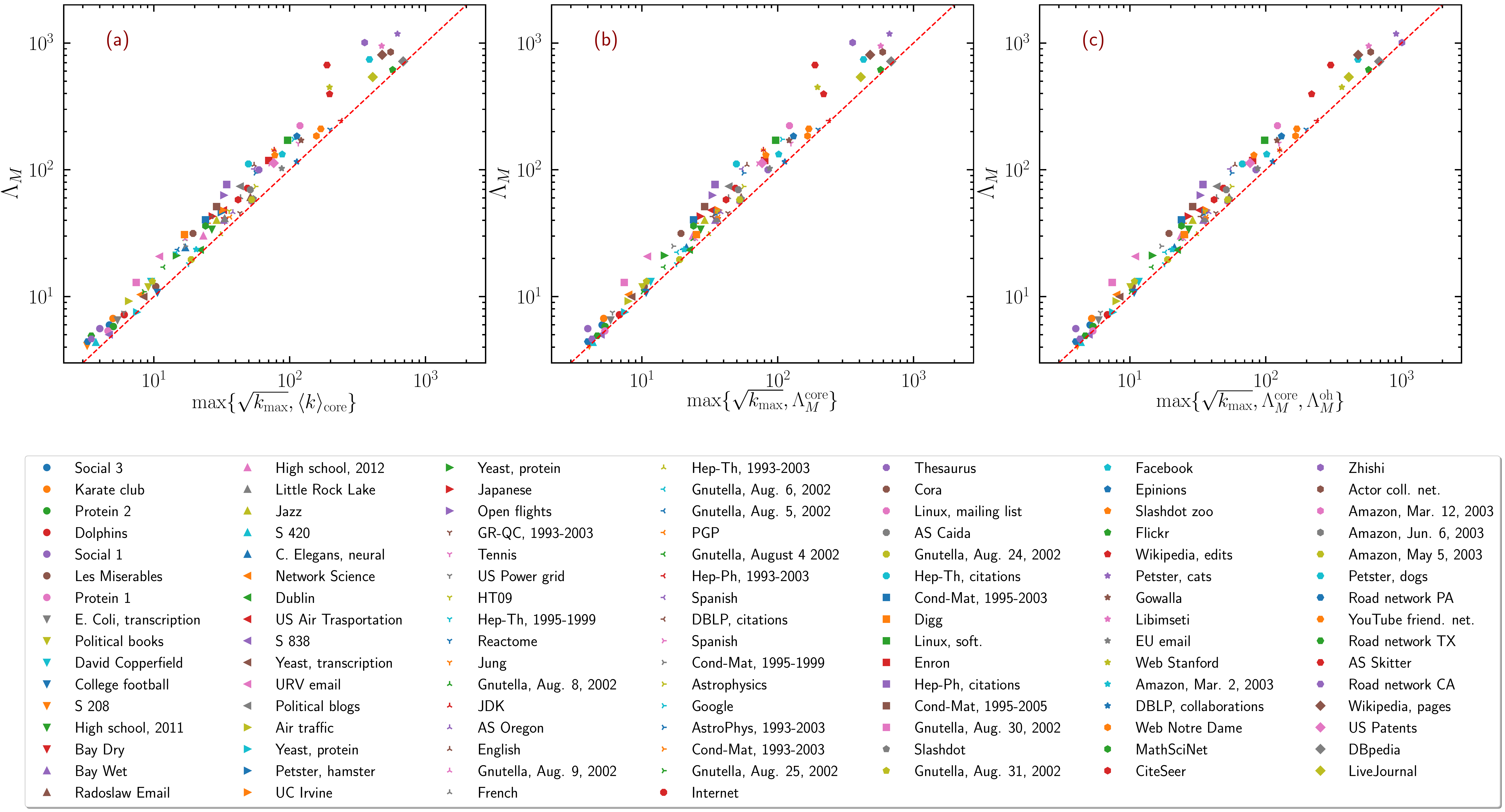}
  \caption{{\bf Check of the theoretical approximations for the
    LEV of the adjacency matrix in real networks.} (a) Value of the LEV
    $\Lambda_M$ of the adjacency matrix as a function of the theoretical
    prediction in Ref.~[23], given as the maximum between
    the square root of the maximum degree and the LEV
    $\Lambda_M^\mathrm{core} $of the maximum $K$-core, approximated by its
    average degree $\av{k}_\mathrm{core}$; (b) same expression, considering
    the LEV of the maximum $K$-core computed numerically; (c) an improved
    version taking into account the LEV $\Lambda_M^\mathrm{oh}$ of the
  maximal set of $n$ overlapping hubs of degree $K$.}
\label{fig:new_figure}
\end{figure}

\begin{figure}[h]
  \centering \includegraphics[width=0.7\columnwidth]{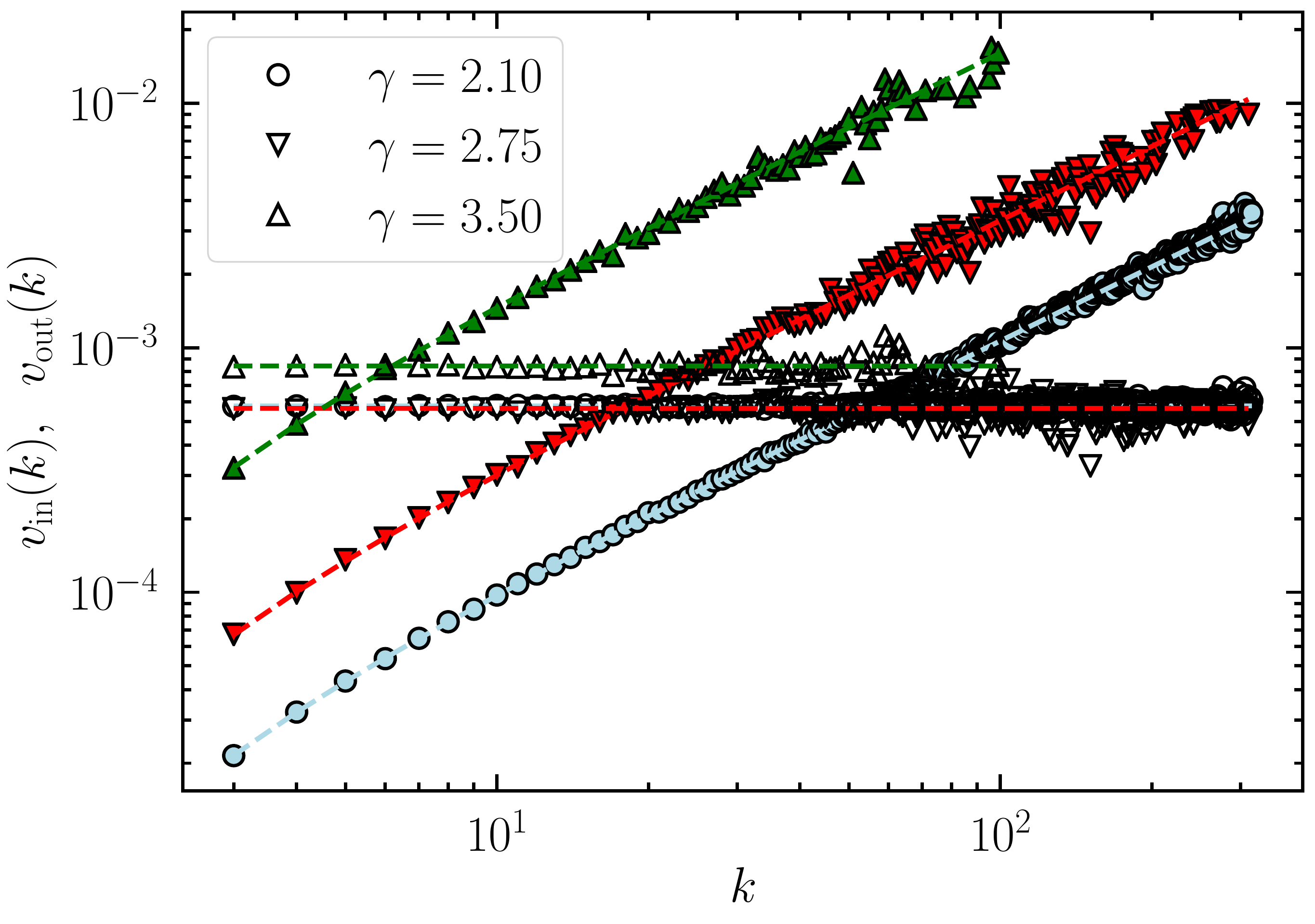}
  \caption{{\bf Behavior of $v_\mathrm{out}(k)$ and $v_\mathrm{in}(k)$.}
    Check of the scaling of $v_\mathrm{out}(k)$ (\change{filled} symbols) and
    $v_\mathrm{in}(k)$  (\change{hollow} symbols) with degree $k$
    in power-law UCM
    networks of size $N=10^5$ and different $\gamma$ exponents. Dashed lines
    denote the theoretical behaviors predicted for  $v_\mathrm{out}(k)$,
    Eq.~(25) and for $v_\mathrm{in}(k)$,
  Eq.~(28).}
  \label{fig:check_v_in_out_UCM}
\end{figure}

\begin{figure}[t]
  \centering
  \includegraphics[width=0.7\columnwidth]{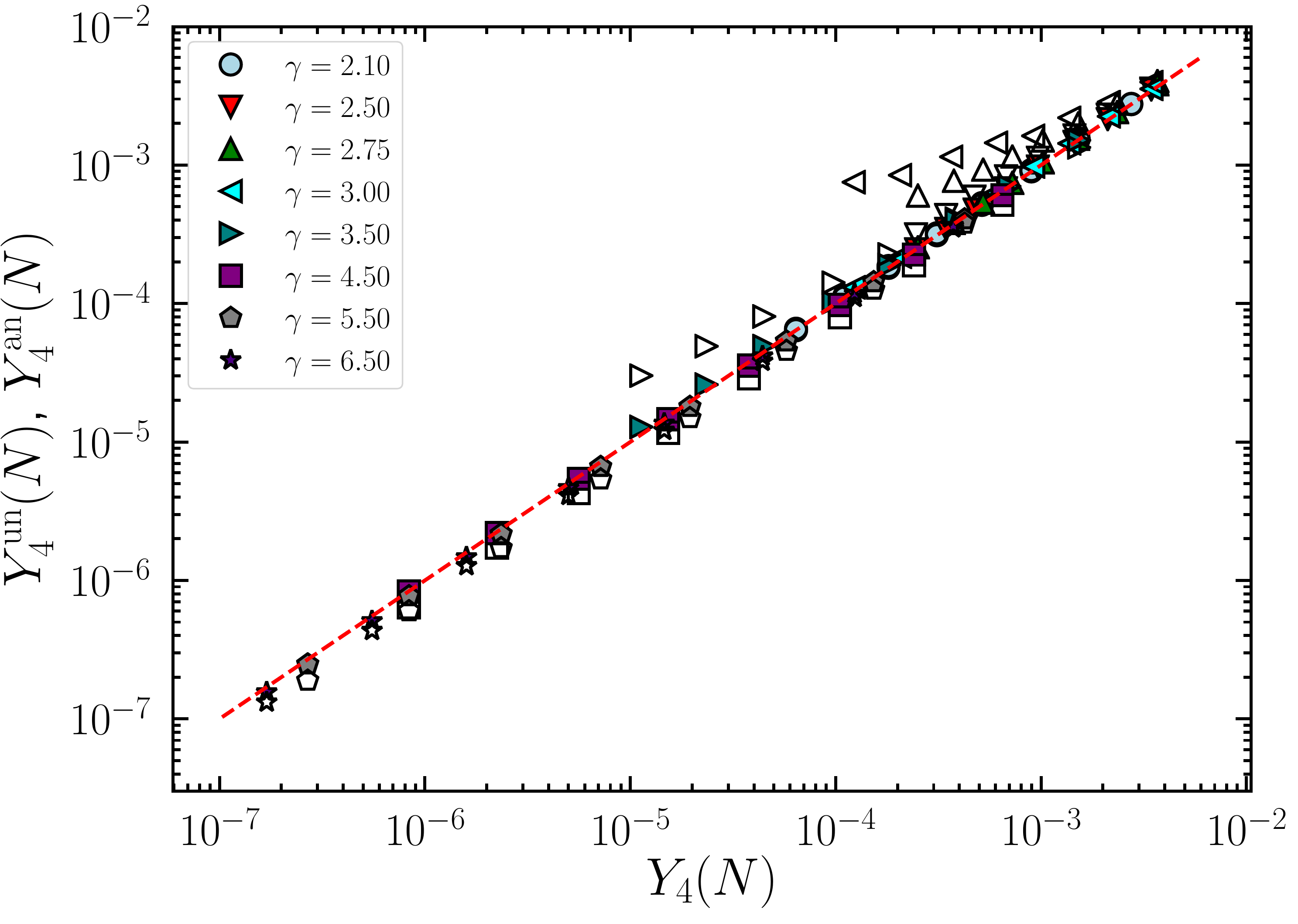}
  \caption{{\bf Localization in synthetic uncorrelated networks.} Inverse
    participation ratio $Y_4(N)$ of the NBC $x_i$ in power-law UCM networks
    with different degree exponent $\gamma$.  We compare, for different
    network sizes,  the results from numerical evaluation  with the
    theoretical prediction $Y_4^\mathrm{un}(N)$ computed from the expression
    $x_i \sim \sum_j A_{ij} (k_j -1)$ (full symbols), and with the
    prediction $Y_4^\mathrm{an}(N)$ from the annealed network approximation
    $x_i \sim k_i$ (hollow symbols). The dashed line represents the behavior
    $y = x$. Simulations results correspond to the average over $25$
    different network realizations of sizes ranging between $N=3000$ and
    $N=10^7$. 
  }
  \label{fig:hashimoto_disparity_prediction}
\end{figure}

\begin{figure}
  \centering \includegraphics[width=0.7\columnwidth]{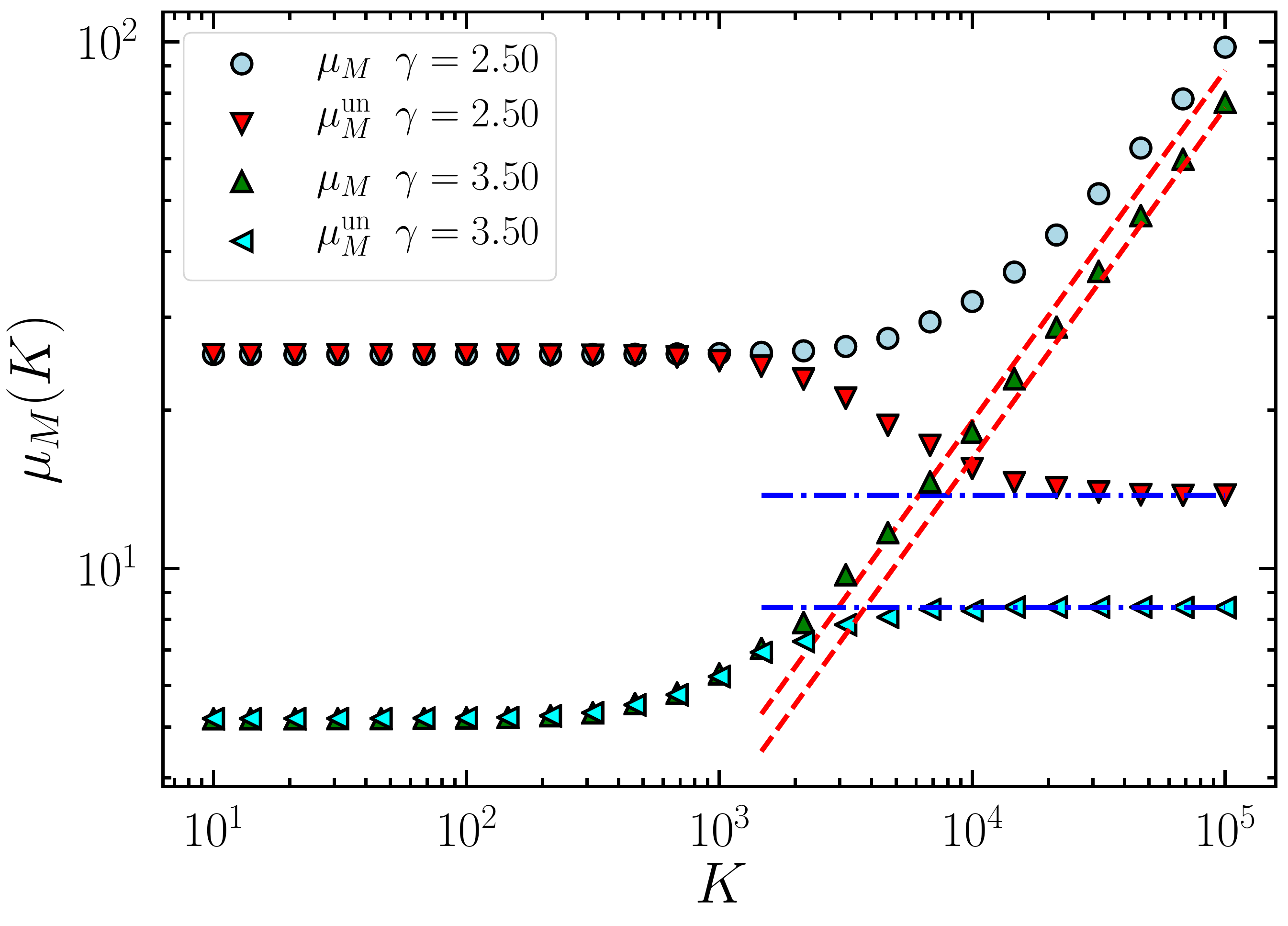}
  \caption{{\bf Effects of the addition of an integrated hub.} Value of
    $\mu_M$ for power-law UCM networks with different degree exponent added
    with an integreated hub of degree $K$.  Dashed lines represent the
    theoretical prediction, $\mu_M^\mathrm{h} =  \left(\frac{\av{k} K(K-1)}{N}
    \right)^{1/3}$. Dot-dashed lines represent the estimation
    $\mu_M^\mathrm{un} \sim 2 \av{k}$, large values of $K$ according to
    the uncorrelated theory, Eq.~(8).
  Network size $N=10^5$.}
  \label{fig:hubeffects}
\end{figure}

\begin{figure}
  \centering \includegraphics[width=0.7\columnwidth]{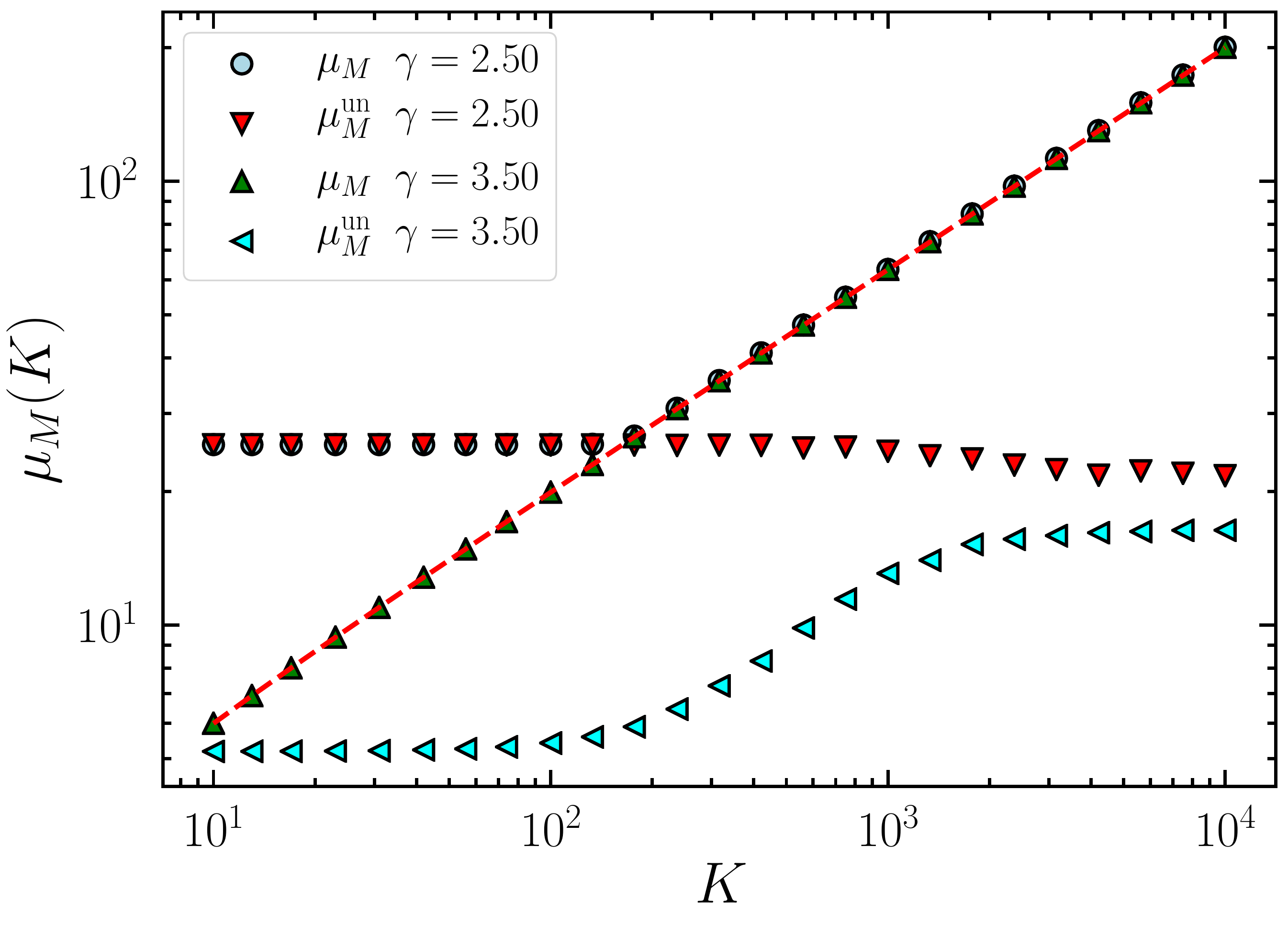}
  \caption{{\bf Effect of the addition of overlapping hubs.} Value of
    $\mu_M$ for power-law UCM networks with different degree exponent, added
    with $n=5$ overlapping hubs of degree $K$.  The dashed line represents
  the theoretical prediction, $\mu_M^\mathrm{oh} =  \left[(n-1)(K-1)
\right]^{1/2}$, independent of $\gamma$.  Network size $N=10^5$.}
  \label{fig:clonedeffects}
\end{figure}

\end{document}